\documentclass[12pt,eqsecnum,preprint]{aastex}

\slugcomment{submitted}

\shorttitle{Flare Observations of the dMe Flare Star EV~Lac}
\shortauthors{Osten et al.}

\begin{document}

\title{ From Radio to X-ray: Flares on the dMe Flare Star EV~Lacertae 
}
\author{Rachel A. Osten\altaffilmark{1}\altaffilmark{a}}
\affil{National Radio Astronomy Observatory \\ 520 Edgemont Road \\ Charlottesville, VA 22903 \\
Electronic Mail: rosten@nrao.edu}
\altaffiltext{1}{Jansky Postdoctoral Fellow}
\altaffiltext{a}{Visiting Astronomer, McDonald Observatory, operated by The University of
Texas at Austin}

\author{Suzanne L. Hawley\altaffilmark{a}} 
\affil{Astronomy Department\\ Box 351580\\ University of Washington\\
 Seattle, WA 98195 \\Electronic Mail: slh@astro.washington.edu}

\author{Joel C. Allred}
\affil{Physics Department \\Box 351560 \\University of Washington\\
 Seattle, WA 98195 \\Electronic Mail: jallred@u.washington.edu}

\author{Christopher M. Johns-Krull\altaffilmark{a}}
\affil{Department of Physics and Astronomy \\ Rice University \\6100 Main Street \\Houston, TX 77005 \\Electronic Mail: cmj@rice.edu}

\author{Christine Roark\altaffilmark{2}}
\affil{University of Iowa}
\altaffiltext{2}{NRAO REU Summer Student}

\begin{abstract}
We present the results of a campaign to observe flares on the M dwarf flare star EV~Lacertae
over the course of two days in 2001 September, utilizing a combination of radio continuum,
optical photometric and spectroscopic, ultraviolet spectroscopic, and X-ray spectroscopic
observations, to characterize the multi-wavelength nature of flares from this active, single late-type
star.  We find flares in every wavelength region in which we observed. 
A large radio flare from the star was observed at both 3.6 and 6 cm, 
and is the most luminous 
example of a gyrosynchrotron flare yet observed on a dMe flare star.
The radio flare can be explained as encompassing a large magnetic volume, comparable to
the stellar disk, and involving trapped electrons which decay over timescales of hours.
Flux enhancements at 6 cm accompanied by highly negatively circularly polarized 
emission ($\pi_{c} \rightarrow -$100\%) imply that a coherent emission mechanism is 
operating in the corona of EV~Lac.
There are numerous optical white-light flares, and yet no signature 
of emission line response from the chromosphere appears.  Two small
ultraviolet enhancements differ in the amount of nonthermal broadening present.
There are numerous X-ray flares occurring throughout the observation, 
and an analysis of undispersed photons and grating events reveals
no evidence for abundance variations.  Higher temperatures are present
during some flares, however the maximum temperature achieved varies from 
flare to flare.  There is no evidence for density variations during any flare intervals.
In the multi-wavelength context,
the start of the intense radio flare is coincident with 
an impulsive optical U-band flare, to within one minute, and yet there is no
signature of an X-ray response.
There are other intervals of time where optical flaring and UV flaring is occurring,
but these cannot be related to the contemporaneous X-ray flaring:
the time-integrated luminosities do not match the instantaneous X-ray flare luminosity, as one 
would expect for the Neupert effect.  We investigate the probability of chance occurrences
of flares from disparate wavelength regions producing temporal coincidences, but find that not all
the flare associations can be explained by a superposition of flares due to a high flaring
rate.
We caution against making causal associations of multi-wavelength flares
based solely on temporal correlations for high flaring rate stars like EV~Lac.
\end{abstract}


\section{Introduction }

The study of extreme coronae, such as on M dwarf flare stars, allows an
investigation into regimes of temperature, density, and activity not available from
spatially detailed studies of our low-activity Sun.  These stars are small, nearly 
(or fully) convective, and have intense magnetic fields covering a large majority of the
stellar disk \citep{jkv1996,saar1994}. 
Despite their small size, dMe flare stars produce close to the maximum amount of coronal
emission any star seems able to maintain --- even in the quiescent state.  Frequent
large outbursts can increase the X-ray luminosity up to $\sim$ 20\% of the total
stellar bolometric luminosity \citep{favataetal2000}.  Flares on these stars are fundamentally
linked to magnetic processes occurring in their outer atmospheres, which involve the entire
atmosphere and produce emission all across the electromagnetic spectrum.  

EV~Lac (dM4.5e) is a young disk population star at a distance of 5.1 pc; its young age comes from kinematics, based on infrared colors \citep{leggett1992}, H$\alpha$ observed to be in emission, 
rapid rotation, and flaring.  It is the
second brightest M dwarf X-ray source seen in the ROSAT All-Sky Survey, with a 
quiescent 0.1--2.4 keV flux of $\sim$ 4$\times$10$^{-11}$ erg cm$^{-2}$ s$^{-1}$ \citep{hunsch1999}.
The quiescent X-ray luminosity is a significant fraction of the total
bolometric luminosity ($\sim$0.2\%). 
Its low mass (0.35 M$_{\odot}$) puts it
near the dividing line for fully convective stars \citep{delfosseetal1998}.  EV~Lac
has very strong, 2.5--3.5 kG magnetic fields covering $>$ 50\% of the stellar surface 
\citep{jkv1996,saar1994},
so coronal phenomena are dominated by magnetic field interactions.

The optical variability of EV~Lac is well known; \citet{abdulaziz1995} detected 25 
flares during 28 hours of five color photometric monitoring.  
\citet{kodaira1976}
saw a flare on EV~Lac with an increase of 5.9 magnitudes in U band.  The star is variable at 
radio wavelengths as well:  \citet{whiteetal1989} determined an upper limit of 0.3 
mJy at 6 cm, while \citet{cdf1988} detected the source at 1.0 mJy. 
\citet{whiteetal1989} also found extremely polarized ($>$ 80\% left circularly polarized) emission at 
20 cm, and \citet{abdulaziz1995} observed bursts of decametric radiation associated with some
of the optical flares reported.  \citet{paa1995} conducted
coordinated IUE and EUVE observations, and recorded a factor of four enhancement in C~IV
emission preceding EUVE flare peaks.  Earlier, \citet{ambruster1986} measured a
decrease in UV line fluxes of a factor of two over 1.5 hours, which they attributed to a 
major mass expulsion episode, similar to solar flare-related eruptive prominences.

EV~Lac has been observed by many previous X-ray satellites 
 and has shown extreme levels of variability.  The types of flares
have varied dramatically:  from extremely short duration events that last a few
minutes with an enhancement of $\sim$ 10 over the quiescent count rate 
\citep{ambruster1994}
to large events that can last up to 24 hours \citep{schmitt1994,sciortinoetal1999} and peak at 300 times the quiescent count rate \citep{favataetal2000}.

Based on its previous record for large and dramatic flare variability, we studied
the multi-wavelength behavior of EV~Lac with observations utilizing the {\it Chandra X-ray
Observatory}, the {\it Hubble Space Telescope}, ground-based optical photometry and spectroscopy,
and ground-based radio interferometry; these observations overlapped during two days 
in September 2001.
The multi-wavelength span of observations allows a probe of the response of different layers of the 
atmosphere to a flare:  radio wavelengths detail the role of nonthermal particles
in flare dynamics; optical continuum
photometric flare variations can be interpreted as emission from a black-body at temperatures
of $\approx$ 10$^{4}$K \citep{hf1992}; optical spectroscopy probes chromospheric emission/absorption lines;
ultraviolet spectroscopy details chromospheric and transition region temperatures; and X-ray
spectroscopy describes thermal coronal plasma.
The focus of this paper is an investigation of flare properties in the regions of
the electromagnetic spectrum involved in the campaign, and the correlation of behaviors at
different wavelengths; a companion paper (Osten et al., in prep.) will discuss the quiescent behavior.

\section{Data Reduction }
\begin{figure}[h]
\begin{center}
\rotatebox{90}{\scalebox{0.5}{\plotone{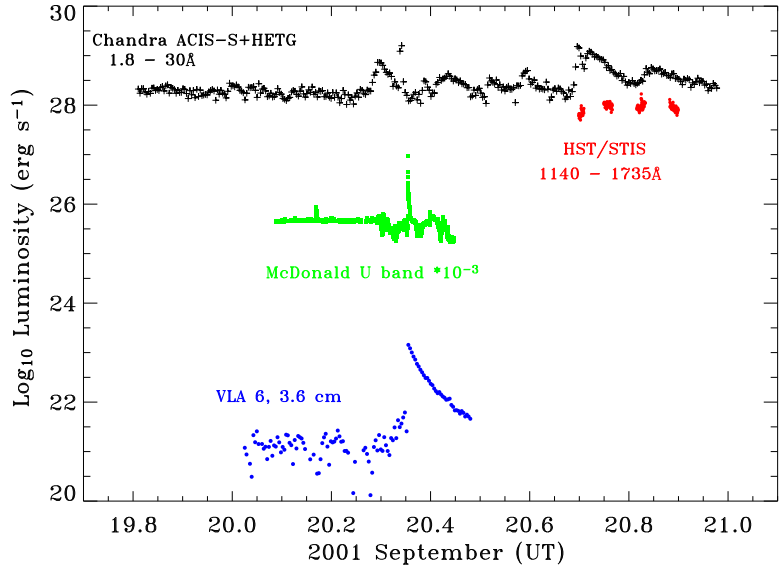}}}
\figcaption[timeline.ps]{ Timeline of observations for the 
2001 September flare campaign; luminosities in each waveband
are plotted to gauge the large-scale variations encountered.  
The {\it Chandra} X-ray pointing, lasting 100ks, formed the
core of the campaign and was supplemented
by ultraviolet, optical, and radio observations.
\label{fig:timeline}}
\end{center}
\end{figure}

A time line of the observations is shown in Figure~\ref{fig:timeline}.  The subsequent sections
describe the individual programs.  

\subsection{VLA Observations}
The star was observed with the NRAO\footnote{The National Radio Astronomy Observatory
is a facility of the National Science Foundation operated under cooperative agreement by 
Associated Universities, Inc.} Very Large Array (VLA).
The phase calibrator was 2255+420, and the flux density calibrator was 0137+331 (3C 48).
On 20 September, the observations were performed
simultaneously in two subarrays at 6 and 3.6 cm (4.9 and 8.4 GHZ, respectively), 
to characterize the multi-frequency
behavior of any flares that might occur during the Chandra observations.  
Processing of the data was done in AIPS.
The field
around EV~Lac is crowded; there is a large radio galaxy $\sim$ 2.7 arcminutes away, and numerous
additional radio sources.  We performed multi-field cleaning, taking advantage of the 
NRAO VLA Sky Survey  
\citep[NVSS][]{nvss} to locate possible bright radio sources in the primary beam and sidelobes.  
After imaging the field, visibilities of sources not identified as EV~Lac were removed,
and the field was re-imaged.  Subsequent analysis utilized the DFTPL program to
extract the time variation of the source flux densities.  The quantities measured are the total
intensity, I, and the amount of circular polarization, $\pi_c$, defined as \\
\begin{equation}
\pi_c (\%)= \frac{V}{I} \times 100 
\end{equation}
where I and V are the Stokes parameters for total intensity and 
circularly polarized flux, respectively. Left circular polarization (LCP) corresponds
to V$<$0; right circular polarization (RCP) corresponds to V$>$ 0.
The correction from International Atomic Time (IAT) to Universal Time Coordinated (UTC) 
was done using the values appropriate for 
2001, $UTC=IAT-31.8 s$ ({\it Astronomical Almanac}).

\subsection{Optical Observations}
Observations were performed at the McDonald Observatory, using the 2.1m Otto Struve telescope
and the 2.7m Harlan Smith telescope.  The 2.1m telescope recorded photoelectric photometry
in the UBVR Johnson-Cousins filters on 18, 19, and 20 September 2001, 
using a two-channel photometer.  EV~Lac was observed
with the primary channel, while the second channel monitored the sky conditions
with a U-band filter.
Integrations in the UBVR filters were 3, 2, 1, and 1 seconds, respectively,  including
0.161 seconds for filter rotation, resulting in a 7 second duty cycle.
Sky measurements were made in the primary channel every 15--20 minutes.  
Extinction and
photometric standard stars \citep{landolt1973,landolt1992} were observed each night.
We concentrate here on the night of 20 September, during which increasing cloud cover
prevented absolute photometric accuracy.
Instead, relative photometry of EV Lac was performed by
dividing the temporal behavior of EV Lac in the primary channel with 
a field star in the second channel
(done in the U filter only).  
We determined the relative variations outside of clouds or flares, normalized
the count rates using this value, and converted
to magnitudes using a quiescent U magnitude of
12.89, as determined by \citet{ac1969}.  
We used the 
zero point magnitude of Vega to convert magnitudes into flux densities.
One large flare and a second smaller enhancement were
detected in the relative photometry, along with smaller variations.  For determination of flare energies, we used the 
U band luminosity of 5.01$\times$10$^{28}$ erg s$^{-1}$ as determined from flux calibrated
data of \citet{gc1969}. Flare energies were determined by using the following formula: \\
\begin{equation}
E = L \times \sum_i \frac{I_{f}-I_{q}}{I_{q}} \Delta t 
\end{equation}
where {\it E} is the net flare energy, {\it L} is the quiescent luminosity in this filter, 
{\it I$_{f}$} is the flare intensity,
{\it I$_{q}$} the quiescent intensity, {\it $\Delta$t} the time step, and
the summation is over the data taken during the flare.

The 2.7m telescope was used in conjunction with the cross dispersed coud\'{e}
echelle spectrograph \citep{tull1995} to obtain high resolution
optical spectroscopy with a time cadence minimum of 55 seconds (45 second exposures
followed by $\sim$ 10 seconds for readout and writing to disk; during cloudy periods 
exposures were lengthened to 60 or 90 seconds).  
The wavelength range was
nominally 3750--9850 \AA\ with small gaps between each of the
58 orders except in the far blue.  Due to the relative faintness of EV~Lac
in the blue, and the short exposure times, the region blueward of H$\delta$
was usually too noisy for reliable measurements.  The data were recorded on a 2048 $\times$
2048 Tektronix CCD which was binned 2 $\times$ 2 to decrease readout time.  As a result of this
binning, the detector undersamples the spectrum.  Spectra of a thorium-argon
lamp are used to determine the wavelength solution of the instrument and to measure the
resolution obtained during the run.  The median FWHM of the 1210 thorium lines
used in the wavelength solution is 1.16 pixels corresponding to a spectral
resolution of $R\approx 55,000$.  The wavelength solution was determined by
fitting a two dimensional polynomial to $m\lambda$ as a function of detector column
number and spectral order $m$.  Spectra themselves are reduced in a standard way using 
software written in the IDL programming language and described by \citet{hinkle2000}.
Briefly, the reductions include cosmic ray removal, dark current and background subtraction, 
and flat-fielding by a normalized exposure of an incandescent lamp.  The continuum
is normalized in each echelle order, enabling measurement of equivalent widths 
and line profile parameters.
Variations in equivalent width of several prominent lines
were determined during the flare.  \\

\subsection{Ultraviolet Observations}
EV~Lac was observed with the {\it Hubble Space Telescope 
Imaging Spectrograph} for 4 orbits on 20 September 2001, 
with a total data accumulation of about 10920 seconds.
The data were obtained with the E140M grating, centered at 1425 \AA\ in TIME-TAG mode
using the FUV-MAMA detector and the 0.2x0.2 arcsecond aperture, to enable investigation
of short time-scale variability at high spectral resolution.  The wavelength range
covered was 1140--1735 \AA\ in 44 orders, with an approximate resolving power of $R=45,800$.

We followed procedures described in \citet{hawleyetal2003} to determine variability and
examine spectral variations.
Spectra were extracted in 60 second intervals to search for large variability; the order location
solution from the pipeline-processed data was used to extract spectra 
and the corresponding wavelength solution was used.
The background and spectral extraction widths were 7 pixels, and
the background was fit by a second order polynomial; the background fit
was then subtracted to obtain a net spectrum. 
Light curves were generated from the 60 second spectra by summing all counts, and subtracting
the sum of the background; Poisson statistics were used to generate 1$\sigma$ error bars.

\subsection{ Chandra Observations}
The campaign centered around a 100 ks {\it Chandra} 
observation, which started 2001 September 19.8, and ended
2001 September 21.0.  The observations were made using the High Energy Transmission
Grating Spectrometer (HETGS) in conjunction with the ACIS-S detector array.
The HETGS provides an image in zeroth-order light of the object as well as through two sets of
gratings:  the Medium-Energy Grating (MEG) covering the wavelength range 1.7--31 \AA\
in first order and the High Energy Grating (HEG) covering the wavelength range 1.2--18.5 \AA\
in first order, with almost twice the spectral resolution.  The observation
was taken in timed exposure mode, for which CCD events are accumulated every 3.24 s 
before being read out.  This sets the minimum time resolution.
{\it Chandra} data are recorded in terrestrial time (TT); we used the correction for 2001
to convert the data to universal time (UTC): $UTC=TT-64.09 s$ ({\it Astronomical Almanac}).

The {\it Chandra} data was processed using 
CIAO, Version 2.2, ``threads'',or
processing recipes, for different aspects of the data reduction. 
We eliminated bad aspect times and confirmed that the observation was not affected
by ACIS ``background flares''.  We applied calibration products: the new ACIS chip 
pixel size (0.0239870 mm from 0.024 mm) and 
focal length (10070.0 mm from 10061.62) were updated in processing the ACIS events.  Events with
energies less than 300 eV and greater than 10,000 eV were excluded, as these are not
well-calibrated; also excluded were events whose pulse-height invariant (PI)
values were 0,1, or 1024 (overflow/underflow values).  Only ASCA grades 0,2,3,4,6
were kept.  The data were resolved into spectral events using region filtering
(rectangles around the MEG and HEG stripes), and a level 2 event file was generated.
One of the CCD chips in the ACIS-S array (S4) suffered increased scatter (streaking)
in the horizontal (CHIPY) direction, apparently caused by a flaw in the serial
readout; a destreak filter was applied to remove these data.

We also extracted spectra corresponding to quiescence, and each flare, from the undispersed
photon events.  The source extraction region was a 3$\times$3 pixel area centered on the
location of 0th order, in which 95\% of the undispersed photons lie.  The background region
was an annulus of 10 pixel radius, with the same center.  
Sensitivity and response matrices were generated for each time slice.
We regrouped the CCD spectra to have a minimum
of 25 photons in each spectral bin; 6 of the 9 flares had sufficient counts that we could perform 
spectral fits, and for a few of the longer duration flares, we attempted further time resolution by
extracting spectra during the rise and decay phases.

We used custom IDL procedures to determine light curve variations and subsequent
spectral analysis.  The level 2 events file was filtered for MEG events falling within
the spectral extraction window ($\pm$0.1 mm in the cross-diffraction
coordinate).  The first 40 ks of the observation were unaffected by flares; we
extracted a quiescent spectrum from this dataset.  Degradation of the ACIS quantum
efficiency (QE) at low energies ($<$ 1 keV) has become a problem, due to the
buildup of a contaminant on the ACIS optical blocking filter.  We updated the sensitivity
files to reflect this change.


\section{Analysis of Individual Datasets\label{analsec}}

\subsection{Radio}
\subsubsection{Large Flare}
\begin{figure}[h]
\begin{center}
\rotatebox{90}{\scalebox{0.5}{\plotone{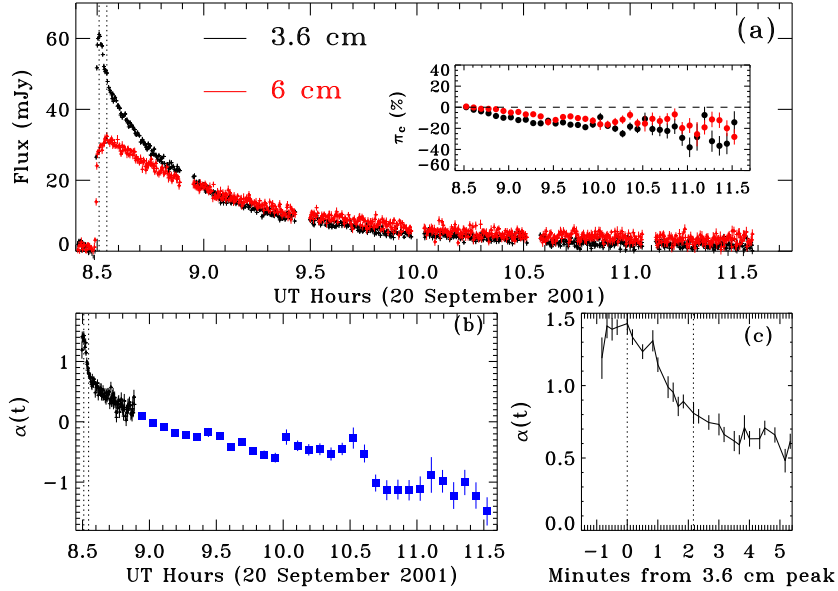}}}
\figcaption[]{{\bf (a)} Variations of total flux density at 3.6 (black) and 6 cm (red)
near the time of the large radio flare on
20 September 2001.
Each tick corresponds to 10 seconds, the temporal resolution available with the VLA.
1$\sigma$ error bars are plotted.  The inset shows variation of percent circularly
polarized radiation in 300 second intervals at both frequencies during the decay of the flare; dashed line indicates
zero polarization.
{\bf (b)} Variation of spectral index $\alpha$, S$_{\nu} \propto \nu^{\alpha}$, between
6 and 3.6 cm ($S_{\nu} \propto \nu^{\alpha}$).  Crosses indicate data sampling of 10 seconds;
blue squares indicate data sampling of 5 minutes (300 seconds). 
1$\sigma$ uncertainty on spectral slope is also plotted.
{\bf (c)} Detailed examination of spectral index variations of the few minutes
around the flare rise and peak: the spectrum appears to flatten for about 40 seconds
before the 3.6 cm peak.
Dotted lines for all panels
indicate times of 3.6 and 6 cm peaks (3.6 cm peaks first).
\label{alpha}}
\end{center}
\end{figure}

Following a long span of small modulations (peaks in intensity $\leq$ 2 mJy)
during the VLA observation on 20 September, 
a large flare occurs toward the end of the 11 hour track 
at both 3.6 and 6 cm in Figure~\ref{alpha}, with a peak in flux of $\sim$ 61 mJy
at 3.6 cm (L$_{R}\sim 1.9\times10^{15}$ erg s$^{-1}$ Hz$^{-1}$). 
Figure~\ref{alpha} also shows the variations in spectral index and circularly
polarized flux during the flare decay.
This flare represents a maximum enhancement of $\sim$150 over the preflare
values; 
previous centimeter-wavelength radio observations had revealed evidence of small
variations \citep{whiteetal1989,cdf1988} on EV~Lac, but the maximum flux density
recorded was only 1.0 mJy. We consider the large outburst recorded at 3.6 and 6 cm
to be a truly unusual event, with a long ($>$ 3 hr) timescale, large (30--60 mJy) peak radio luminosity,
yet small ($<$20\%) circular polarization. 

The spectral indices start out very steep, 
and decline approximately linearly.  
Panel (c) shows a close-up of
the spectral index behavior:  for $\approx$ 40 seconds before the 3.6 cm peak the 
spectral index is roughly constant, at $\alpha \sim 1.4$.
Such large spectral indices during the initial phases of the flare indicate large
optical depths in the flaring plasma; both frequencies are likely optically thick.
At late stages in the flare
the spectral index has reached a value of $\alpha \sim -1.6$ and the radiation 
at 3.6 cm is likely optically
thin. 
The decline in flux is not characterized by a 
single exponential decay at either frequency; instead, the timescale $\tau$ over which the radiation 
decays exponentially (flux $\propto \exp(-(t-t_{0})/\tau)$; t$_{0}$ is time of peak flux)
progressively lengthens during the flare decay.  
Table~\ref{tbl:raddecay}
lists the exponential decay timescales at both frequencies during various stages of the 
flare decay.  
The 3.6 cm flux decays at a 
faster rate (smaller decay timescale) than the 6 cm flux, but the increase in the 
timescale as the decay proceeds occurs at both frequencies.

The interpretation of this flare is complicated by optical depth effects early in 
the flare, when electron acceleration, injection into a flare loop, and particle trapping 
are most likely to occur.  In contrast, the late stages of the flare 
 appear to be consistent with optically thin
emission at the higher frequency (3.6 cm, or 8.4 GHz).
We interpret the variations at both frequencies under the assumption
that the emission mechanism is gyrosynchrotron radiation from a nonthermal population of mildly relativistic
electrons, whose distribution with energy is described by $N(E) \propto E^{-\delta}$ 
\citep[see][]{dulk1985}. 
The spectral indices obtained during the flare rise, peak and initial
decay, however, are not consistent with the range expected from a homogeneous optically thick source
\citep[2.5--3.1 for $\delta$ between 2 and 7;][]{dulk1985}, and therefore it is probable that multiple sources with different magnetic field strengths (perhaps an arcade of loops)
are contributing to the observed radio emission.
It is difficult to separate
the temporal behavior of the emission (number and distribution of accelerated 
electrons, changing magnetic
field strengths) from temporal change of the source size; both are conflated
in measurements of total flux density.

\begin{figure}
\begin{center}
\scalebox{0.6}{\plotone{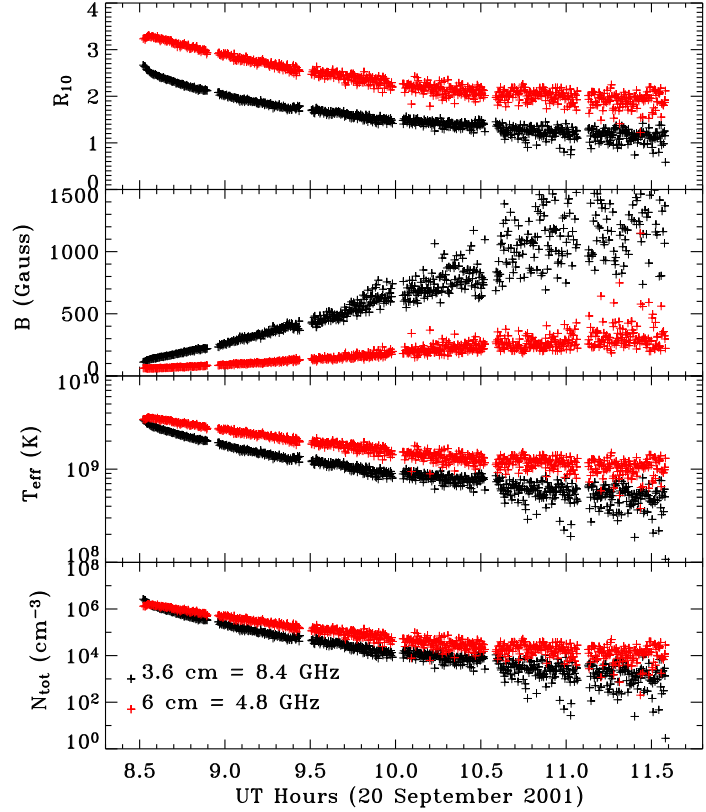}}
\figcaption[]{Model parameters derived during the decay of a large radio flare, under the assumption of optically thick gyrosynchrotron emission from an inhomogeneous, dipole magnetic
field configuration.  R$_{10}$ is the source size in 10$^{10}$ cm, B the magnetic field
strength of the radio-emitting source, T$_{\rm eff}$ the effective temperature of the 
plasma, and N$_{\rm tot}$ the total number density of nonthermal electrons above a
cutoff value of 10 keV; see text for details.  \label{radmod}}
\end{center}
\end{figure}

We investigated some of the key parameters which control the amount of radio emission
radiated during the flare.  
We assume that the emission at both 3.6 and 6 cm is optically
thick, and originates from an inhomogeneous magnetic field with a large dipole configuration,
or $B(r)=B_{0} (r/r_{0})^{-3}$, where $B(r)$ describes the magnetic field strength distribution
as a function of radius.
The assumption of optically thick emission allows us to equate the
brightness temperature with the effective temperature of the plasma, and implies that the 
emission is coming from a layer where the observed frequency is the peak 
frequency, i.e., where $\tau=1$.  We assume that the distribution of nonthermal
electrons remains a power-law with fixed index $\delta=3$ through the flare decay.
Using the analytic expressions in \citet{dulk1985}, the 
flux density can be expressed as  \\
\begin{equation}
S_{\nu} = 2k_{B} T_{B} \Omega \frac{\nu^2}{c^2} = 2k_{B} T_{\rm eff} \frac{\nu^2}{c^2}
\frac{\pi r_{s}^{2}}{D^{2}}
\end{equation}
where S$_{\nu}$ is the flux density, k$_{B}$ is Boltzmann's constant, T$_{B}$ is the brightness temperature,
T$_{\rm eff}$ is the effective temperature, $\nu$ is the observing frequency,
$\Omega$ is the solid angle subtended by the source, $r_{s}$ is the source size
and $D$ is the distance to the source.  Adding in equation 37 of \citet{dulk1985} describing
the dependence of effective temperature on $\delta$, $\nu$, B, and angle $\theta$ 
between magnetic field and line of sight, we solve for the changing 
source size as a function of time at the two radio frequencies, 
assuming that $r_{0}$ is the stellar radius, $B_{0}$ is
$\approx$ 3000 G, $\delta=3$, $\theta$=60$^{\circ}$.  
The small spectral indices in quiescence indicate a shallow
number distribution with energy, compatible with
$\delta=$2 or 3, and we choose $\delta=$3. This is also compatible with
previous models for nonthermal stellar radio emission, which find that flat
spectral energy distributions are better able to reproduce observed
spectral energy distributions \citep{whiteetal1989b}.
The dependence on angle $\theta$ is not steep, and we choose an angle $\theta$ 
of 60$^{\circ}$ as representative.
Figure~\ref{radmod} plots the
variation of the source size, magnetic field strength in the source, and effective temperature
for the two radio frequencies during the decay of the large flare.  Once the magnetic field
strength in the source is known, the total number density of nonthermal electrons above a cutoff
energy E$_{0}$, can be calculated, using equation (39) of \citet{dulk1985} (in these
calculations E$_{0}$=10 keV is assumed); this quantity is also plotted in Figure~\ref{radmod}.
Note that the source sizes are larger than or comparable to the 
stellar radius R$_{\star}\approx$2 r$_{10}$ cm.  The minimum magnetic field strengths
at the flare peak are 60 and 110 Gauss, at 6 and 3.6 cm respectively, and maximum
effective temperatures $\sim$3.4$\times$10$^{9}$K.  
The higher frequency indicates a larger field strength at all times due to the larger gyrolayers
being sampled.
The magnetic field strength increases with time during the flare decay, a consequence of the decreasing
source size, as the location of optical depth unity moves closer to the stellar surface.
These values are consistent
with values previously determined from e.g. VLBA studies of M dwarf coronae \citep{benz1998}.

The major energy loss mechanisms for accelerated electrons during solar flares 
are Coulomb collisions from electrons caught in a magnetic trap, and
precipitation from the trap into the chromosphere via scattering into a loss cone 
\citep[so-called trap plus precipitation
models;][]{melrosebrown1976}.  
Coulomb collisions preferentially deplete low energy electrons compared to high energy electrons;
the timescale for radiative loss goes as E$^{3/2}$, and the observed flare variation at each frequency
is qualitatively consistent with the hypothesis that Coulomb collisions 
dominate during the decay phase,
as the timescales grow larger during the decay.
The collisional deflection time  is approximated in the weak diffusion limit by \\
\begin{equation} \label{eqn:taud}
\tau^{\rm defl}= 9.5\times10^{7} \big( \frac{E_{\rm keV}^{3/2}}{n_{e}} \big) \big( \frac{20}{ln[8\times 10^{6} T_{e}/n_{e}^{1/2}]} \big) \;\;\; s
\end{equation}
\citep{benz2002}, where {\it n$_{e}$} and {\it T$_{e}$} are the ambient trap particle density and temperature,
respectively; and {\it E$_{keV}$} is the energy of the accelerated particle in units of keV.
By analogy with solar flare behavior, 
the observed decay timescales for the flare on EV~Lac listed in Table~\ref{tbl:raddecay} 
are consistent with very energetic electrons, or a 
remarkably low trap density, or a combination of the two.  A trap density of $n_{e} \sim 10^{8}$
cm$^{-3}$ and thermal electron temperature of T$_{e} \sim$10$^{7}$K are consistent 
with the observed decay times and with 
particles of $\approx$ 100--400 keV in energy.
During the flare decay, then, one expects that the low energy particles will lose energy
and precipitate from the trap, resulting in a hardening of the electron distribution 
with energy.  The toy model for magnetic field inhomogeneities described above
assumes that the number distribution with energy has a fixed index, and only the 
total number density above a cutoff energy E$_{0}$ decreases.  The depletion of low energy
electrons from the trap could be simulated in the model by a corresponding 
increase in the cutoff energy
in the electron density distribution with energy.  The ratio of nonthermal electron density
to thermal electron density in the trap implied by the model is very high, $\sim$10$^{-2}$
at the beginning of the flare decay.  

\subsubsection{Small Flares}
\begin{figure}[h]
\begin{center}
\scalebox{0.5}{\plotone{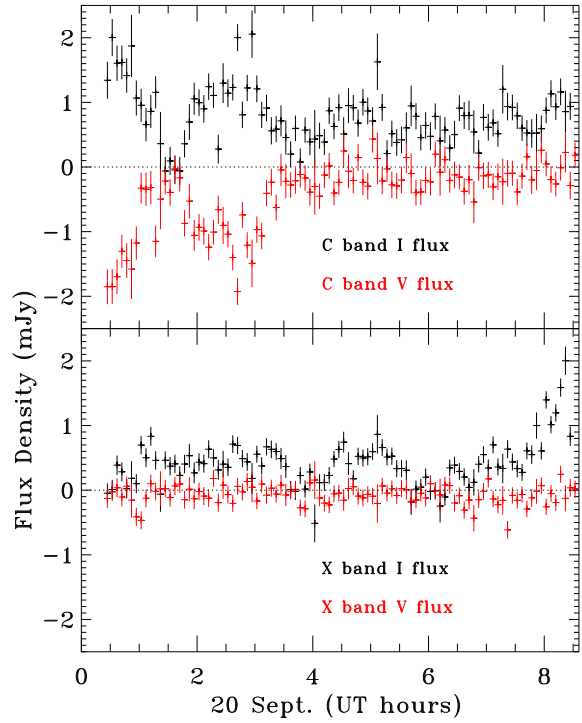}}
\figcaption[]{{\bf (top)} Variation of total intensity (I) and circularly polarized
flux (V) at C band (6 cm) during the first 8 hours of the VLA observation; data has
been binned to 300 second intervals.  
Two events characterized by enhanced I and V fluxes are noticeable.
{\bf (bottom)} Variations of total intensity (I) and circularly polarized flux
(V) at X band (3.6 cm) during the first 8 hours of the VLA observation.  
The variability seen at C band is not seen at this higher frequency; binning is same as 
above. 
\label{cbandflare} }
\end{center}
\end{figure}

Several smaller amplitude flares are also evident in the data, particularly
at 6 cm (Figure~\ref{cbandflare}), accompanied by large amounts of 
circularly polarized flux.
Computations of  ``preflare'' values of average flux
and polarization were determined by limiting the time range to 03:30:00 -- 08:25:00, to avoid
any of this smaller scale variability (although it is still apparent at 3.6 cm).  
The average preflare fluxes at 3.6 and 6 cm
are 0.41$\pm$0.02 and 0.65$\pm$0.03 mJy, respectively, and the average circular polarization
(average V flux divided by average I flux)
at 3.6 and 6 cm is $-$22$\pm$5\% and $-$22$\pm$5\%, respectively.
There are two events at 6 cm which reach 100\% circular polarization, 
with no apparent correspondence at 3.6 cm. These events occur in
both intermediate frequencies used in the 6 cm observations, spanning 
4810--4910 MHz.  The lack of corresponding variations at 3.6 cm constrains the bandwidth
of the event to be 100 MHz $\leq \Delta \nu \leq$ 3500 MHz.
The peak amplitudes of a few mJy are rather small,
implying that the peak radio luminosity is 6.2$\times$10$^{13}$ erg s$^{-1}$ Hz$^{-1}$.
This behavior is very reminiscent of radio bursts with large amounts of circular polarization
seen on other active M dwarfs, which have accompanying large brightness temperatures (generally 
$>$10$^{13}$K) and short durations, although an event on the dMe star Proxima Centauri
reported by \citet{slee2003} had 100\% right circular polarization lasting for days.

The two highly polarized events early in the observation are relatively long-lasting,
having durations of $\sim$ 1 hour each.  
The large amounts of circular polarization argue for a relatively localized phenomenon, due to the
large brightness temperatures needed ($\geq$10$^{13}$K) to generate extreme circular polarization
levels (usually a coherent process is needed); 
the source size can be calculated as \\
\begin{equation}
r_{10}=\frac{d_{pc}}{8.7} \left( \frac{5 GHz}{\nu} \right) \left( \frac{S_{\nu}}{100 mJy} \right)^{1/2}
\left( \frac{4.8\times10^{10}}{T_{b}} \right)^{1/2}
\end{equation}
where r$_{10}$ is the source size in 10$^{10}$ cm, d$_{pc}$ is the distance to the source in parsecs,
$\nu$ is the observing frequency, S$_{\nu}$ the flux density, and T$_{b}$ the brightness
temperature; using a lower limit of 10$^{13}$K for the brightness temperature requires r$_{10}$
to be less than 6$\times$10$^{-3}$, or about 0.1\% of the stellar diameter.

\subsection{Optical }
Two flares are evident in the U band light curve depicted in Figure~\ref{fig:timeline}.  
The first is an
enhancement of only 0.68 magnitudes, lasting $\sim$ 268 seconds (peak at 20 September 04:04:34 UT), 
while the other has a peak of 3.3 magnitudes 
above the quiescent value and lasts
for about 420 seconds (peak at 20 September 08:29:31 UT).  
Using the formula described in \S2.2, we estimate the flare energies to be 2.6$\times$10$^{30}$ 
ergs and 5.2$\times$10$^{31}$ ergs
for the two flares, respectively.
The existence of clouds prior to and during the larger flare makes precise
determination of the flux variations difficult, but there appear to
be smaller enhancements between 06:00 and 08:00 UT.  
We are able to determine
relative flux variations in the U-band filter only because the second photometer channel also had a
U-band filter in it monitoring another star; the variations in BVR cannot be disentangled from cloud variations, but we do clearly see the larger flare in all the colors.

\begin{figure}[h]
\begin{center}
\scalebox{0.5}{\plotone{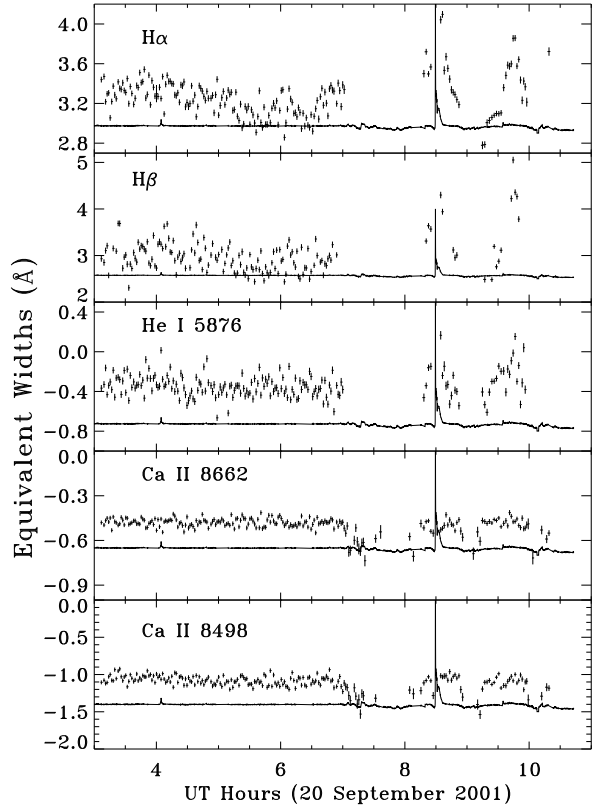}}
\figcaption[]{Variation of equivalent widths of H$\alpha$, H$\beta$,
He~I$\lambda$ 5876, and Ca~II $\lambda\lambda$8662,8498.  
Overplotted is the U band light curve.
\label{opticaldata} }
\end{center}
\end{figure}

We also investigated the variations of equivalent widths of chromospheric lines 
--- the data are displayed
in Figure~\ref{opticaldata}.  
The presence of clouds reduced the flux, thereby
making continuum normalization difficult (due to the decreased count rate) and adding systematic
error to the resultant equivalent width calculations; for this reason we excluded spectral time slices
based on low signal-to-noise of the spectral order (generally SNR $<$ 10).
There is some evidence for variation of H$\alpha$ during the flare decay,
however due to the variable cloud coverage it is not possible to quantify accurately how much
variation took place.  We note that previous studies of EV~Lac \citep{bgs2001quiet,bgs2001flare}
determined a range of H$\alpha$ equivalent widths of 3.5--5.3 \AA,
and H$\beta$ equivalent widths of 4.4--5.0 \AA\ during quiescent periods;
during ``active states'' the range of equivalent widths they observed was 4.7--10.6 \AA\
and 10.2--20.9 \AA, respectively.  As shown in Figure~\ref{opticaldata}, our equivalent width measurements
span $\sim$3--4 \AA.
There was an impulsive U-band flare during our
observations around 08:29:31 UT, and we observe some variation of equivalent width
in H$\alpha$ and H$\beta$,
but the existence of similarly large variations one hour later
(when no photometric flare indicators are present) make the flare interpretation problematic.
\citet{hawleyetal2003} noted from studies of flares on the M dwarf flare star AD~Leo
that optical chromospheric lines have a long gradual phase, compared to the impulsive
continuum variations, and if such events are common to flare stars, we should have had
an increased probability of observing enhanced equivalent widths during the flare.

\subsection{Ultraviolet}
Two small enhancements can be seen in the UV light curves in Figure~\ref{fig:timeline}, in orbits 1 
and 3 of the four HST orbits in which data were taken. The peaks were reached 
at 16:53:46 and 19:48:00, 
represent enhancements of $\sim$ 1.5 (1.8) times the nearby count rates, and last for
only about 4 (5) minutes, respectively.
The total UV energy released during the two flares was estimated by subtracting from 
each the quiescent spectrum, and integrating from 1270--1750\AA; the first flare radiated
1.4 $\times$ 10$^{29}$ erg, the second 5.0 $\times$ 10$^{29}$ erg.
We attempted to determine spectral variations during the events.
Previous analyses of spectral characteristics of ultraviolet flares on active stars revealed a 
variety of phenomena, from downflows with velocities up to 1800 km s$^{-1}$ \citep{bwb1992} 
to broadening of Si~IV line profiles with widths of up to 500 km s$^{-1}$ \citep{linskywood1994},
to blue-ward enhancement of continuum emission during flares \citep{robinson2001}.
However, due to the low enhancement above the non-varying count rate, and short time duration
of these events, the signal-to-noise of the 60-second flare spectra was not high enough
to determine any of these reliably.

\begin{figure}[h]
\begin{center}
\scalebox{0.5}{\plotone{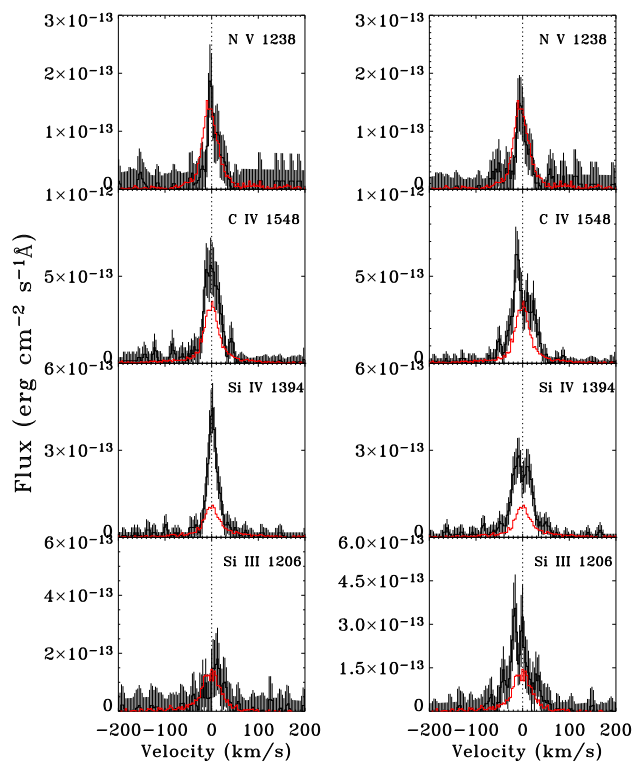}}
\figcaption[]{{\bf (left)} Variation of strong UV emission lines during a small flare in the first
orbit of HST observations.  Overplotted in red is the line profile corresponding to
quiescent times.  This flare lasted 4 minutes, with a peak enhancement of
about 1.5 times the neighboring count rates.  
{\bf (right)} Same as left panel, for a small flare in the third orbit of HST
observations.  This flare lasted $\sim$ 5 minutes, with a peak enhancement of about 1.8 times
the neighboring count rates.  The formation temperatures of the lines decrease downward. 
Flare spectra have been smoothed over two wavelength bins.
\label{uvcompare}}
\end{center}
\end{figure}

We next co-added all the 60 second spectra which 
occurred during flares, to investigate gross flare-quiescence differences and
flare-flare differences.  
An examination of the strongest features during the two flares, in comparison with
quiescent line profile shapes, is given in Figure~\ref{uvcompare}.
Our fitting
routine \citep[based on][]{hawleyetal2003} estimates the surrounding continuum level, and performs single and double-Gaussian line-profile fitting
to the brightest lines in the flare spectra; the results are listed in Table~\ref{tbl:uvtbl}.

A feature common to transition region emission line profiles from active stars is extended wings
in the line profile \citep{woodetal1996} which are fit better by two Gaussian profiles than a single
Gaussian.  For the quiescent spectrum, this is the case; however, the flare spectral profiles did not
show any statistically better fit using two Gaussians than one (due partly to signal-to-noise constraints)
and therefore we fit only a single Gaussian profile to the features in the flare spectra.  
The flare profiles analyzed by \citet{linskywood1994} on the dMe flare star AU~Mic, 
while showing remarkable amounts of broadening,
also were consistent with single Gaussian shapes.
None of the lines analyzed here shows conclusive evidence for bulk plasma motions during either flare.  
Only for the narrow component of the quiescent spectrum does there appear to be a
deviation from zero, and it appears to be 
temperature-dependent, with the magnitude of the blueshifts increasing
towards higher
temperatures.

We investigated evidence for nonthermal motions in the
line profiles, presumed to have a Gaussian distribution, and present in the line profile 
widths above the expected
thermal widths of the emission lines.  The turbulent velocities deduced for the two 
flare segments are comparable 
to that deduced from the narrow component of the quiescent lines, and appears to be
temperature-dependent, in that temperatures below 10$^{5}$K show evidence of excess
line widths, while the highest temperature line examined (N~V $\lambda$ 1238) shows
no turbulent broadening.

The second flare has line widths almost 
twice those of the first flare,
and the narrow component of the quiescent features, although the broad component to the quiescent features
is still largest.  The flux enhancements of the two flares are similar for the high temperature lines
probed in Table~\ref{tbl:uvtbl}, with the exception of Si~III $\lambda$ 1206, for which the second flare
has three times the flux of the first.

\subsection{X-ray}
\begin{figure}[h]
\begin{center}
\rotatebox{90}{\scalebox{0.5}{\plotone{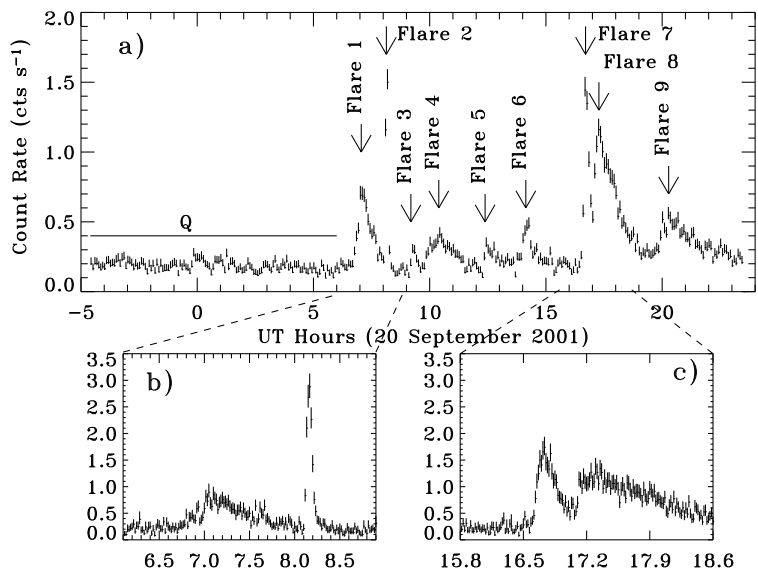}}}
\figcaption[]{(a) Light curve of {\it Chandra MEG} data.  Each tick mark represents an average
over 300 seconds of data; error bars span $\pm$1 $\sigma$ uncertainties of the count rate.  At least nine large (factor of $>$ 2 enhancement over quiescent)
flares are noticeable. Panels (b) and (c) detail two impulsive flares; the time binning is 
60 seconds.  Flare 2 lasts at most 8 minutes, with peak count rate $\sim$15 times the
quiescent count rate, while flare 7 lasts 25 minutes with maximum enhancement 8.5.
\label{chanlc}}
\end{center}
\end{figure}

Numerous X-ray flares are evident in the light curve of Figure~\ref{chanlc}, shown binned
at 300 seconds.  At least nine flares with a peak enhancement of $>$ 60\% over the
surrounding count rate are noticeable, and they are clumped in time, occurring during the
last 60 ks of the observation. The first 40 ks appear to be characterized by lower-level
modulations which may be evidence of smaller amplitude variability.  
The average count rate in the first 40 ks is 0.17$\pm$0.03 ct s$^{-1}$, and
the X-ray luminosity during this time from dispersed spectra (1.8--26\AA) is 1.8$\times$10$^{28}$
erg s$^{-1}$.
We estimate the 
lower limit of the X-ray
flaring rate to be 0.32 flares hr$^{-1}$, using the nine individual flares noted in 
Figure~\ref{chanlc}.
Several flares appeared extremely impulsive, and we investigated variations by extracting a
light curve with a finer
time grid of 60 seconds.  Panels (b) and (c) of Figure~\ref{chanlc} detail flares with the largest
enhancements above the nonflaring count rate.  We determined the time and count rate corresponding
to the flare peak using the 60 second light curve for each of the nine flares, and
extracted spectra (both grating and 0th order) corresponding to each of these flares.
The number of X-ray photons extracted from each flare was not enough to assemble a
dispersed MEG spectrum with good enough statistics to perform an investigation of
flare-to-flare variations.  We used the MEG data from each flare to estimate the excess
energy radiated during each flare, and these numbers, together with count rates and durations,
are tabulated in
Table~\ref{tbl:chanflare}.  

\subsubsection{Analysis of Zeroth Order Spectra}
The location of the undispersed light falls on ACIS chip S3, which has FWHM energy resolution 100
-- 200 eV between 1 keV and 8 keV; 
in addition, the spectrum suffers from pileup effects, due to the detection of two soft photons as
one hard photon with the combined energy of the two actual photons.  This can cause an
artificial hardening of the apparent X-ray spectrum.  Since 95\% of the 0th order photons
fall within a central 3$\times$3 pixel area, and EV~Lac has a low count rate ($\sim$0.17 counts s$^{-1}$
during quiescence), it is not feasible to examine the 0th order time-resolved spectra outside
of the piled-up regions.  A correction for the effect of pile-up has been implemented in 
Sherpa and XSPEC using the model of \citet{davis2001} to correct for grade migration.
We model the zeroth order spectra using as multiplicative models the pileup and interstellar
absorption, and additive models two discrete temperature APEC models within XSPEC
\citep[using the abundances of][]{ag1989}.
Although \citet{sciortinoetal1999} used an interstellar hydrogen column density
of $\sim$10$^{19}$ cm$^{-2}$ in analysis of ROSAT and BeppoSAX observations, we kept
N$_{H}$ fixed to 10$^{18}$ cm$^{-2}$ in analyzing the ACIS spectra, as {\it EUVE}
spectra constrain it to be lower than a few times 10$^{18}$ cm$^{-2}$ (Osten et al, in prep.).

\begin{figure}[htbp]
\begin{center}
\rotatebox{90}{\scalebox{0.6}{\plotone{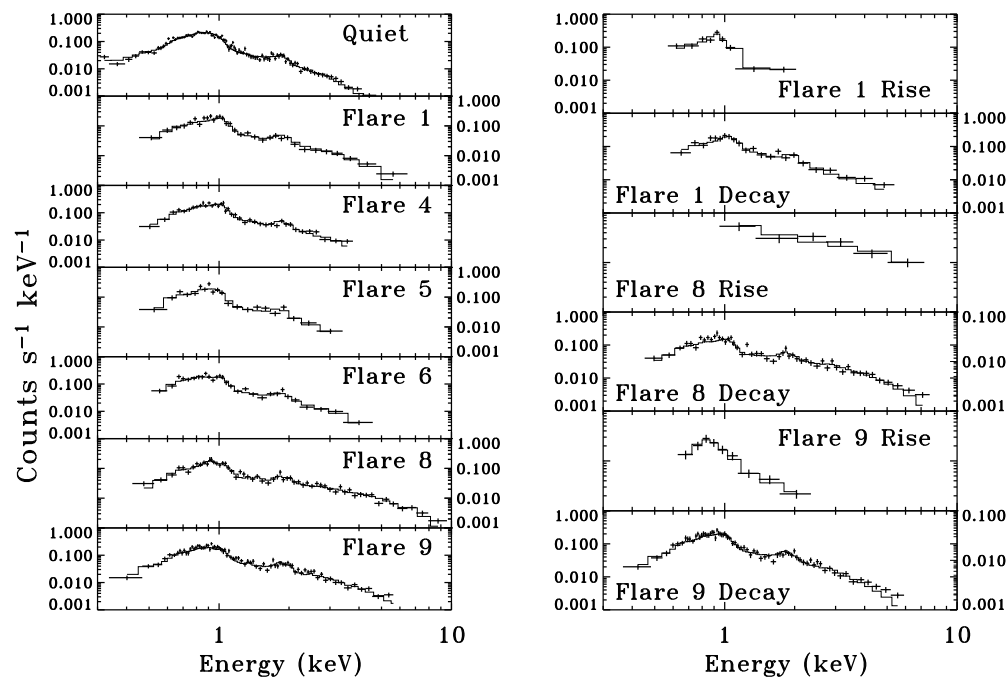}}}
\figcaption[]{ACIS 0th order spectra of differing temporal regions (defined in Figure~\ref{chanlc}).
Crosses indicate data points and errors; histograms illustrate model fit to spectra.  Model
parameters are indicated in Table~\ref{tbl:0thorder}.  Left panels illustrate spectra accumulated
in broad activity bins, corresponding to individual flares (or quasi-steady conditions); right
panels attempt to differentiate changing spectral conditions during individual flares with 
sufficient
durations. \label{0orderspec}}
\end{center}
\end{figure}

Table~\ref{tbl:0thorder} lists the spectral quantities derived from quiescence, as well
as six of the nine individual flares, and Figure~\ref{0orderspec} 
displays the spectra and model fits.  
The lowest temperature component in all cases
is consistent with a value of about 4 MK, while the high temperature component varies from
9 to 30 MK.  
Flare 8 is the interval which shows the hottest plasma temperature.
The global abundance, Z, appears to be low in all
spectra, but shows no significant variations; 
this is limited, however, by the large error bars for
some flare spectra, which prevent determining variations of factors of three or less.
For flares 8 and 9, no convergence could be found for a simple global metallicity scaling,
and instead we decoupled the abundances and solved for O, Ne, Mg, and Fe separately.  
There is again no evidence for 
abundance variations, within the large error bars for the derived values.

For the three longest lasting flares (1, 8 and 9) we extracted spectra corresponding
to the rise and decay phase, separately, to investigate possible intra-flare plasma changes. 
The results are also listed in Table~\ref{tbl:0thorder} and displayed in the right panel
of Figure~\ref{0orderspec}.  There were only significant counts during the decay phases,
and the flare decay spectra show a consistent pattern, being well-described
by a two-temperature fit, the lower temperature corresponding to values derived
from quiescent intervals, and the higher temperature between 11 and 20 MK;  the global 
abundance is similarly low in all three cases.
During the decay of Flare 1 we found it necessary to fix the 
temperature and emission measure of the cooler component, allowing the higher temperature
and emission measure, and global metallicity, to vary.
There is no evidence of dramatic
abundance variations, as have been found from studies of other stellar flares.  
For Flare 1,
there is no significant difference between the temperatures derived from the entire flare and
those derived from the decay, suggesting that the rise phase of the flare was not hot enough
(or did not contain enough material) to affect the integrated flare spectrum.  
For Flares 8
and 9, however, the hotter temperature component during the decay phase is
smaller than obtained from the integrated flare spectra.  

We attempted to fit simplistic, one temperature models with pileup
to the few counts in the rise phase spectra.
The spectra during the rise phases are remarkably different from each
other (and from the decay spectra); 
the rise phase of flare 8 implies very hot temperatures, while the rise of flare 1
shows no significant departure from the quiescent temperatures, and the rise of flare 9 
shows a modest (factor of two) increase.  The abundances of the rise phase spectra had
to be treated differently in each case, in order to get a convergent fit.  
For the rise phase spectra of flare 1, we decoupled the Ne abundance from the others,
finding somewhat enhanced Ne; however, the large error bars swamp this. 
For the rise phase spectra of flare 8, we fixed the abundance at solar due to
the low number of spectral bins.  For the rise phase spectra of flare 9, we allowed
the global abundance to vary freely, finding generally low values.  For these spectra 
with poor spectral resolution and S/N, it is difficult to conclude whether these 
fits reveal evidence for abundance anomalies.

\subsubsection{Analysis of Grating Spectra}
The number of X-ray photons extracted from each flare was not enough to assemble a dispersed
MEG spectrum with good enough statistics to perform an investigation of flare-to-flare variations.
Instead, we co-added time intervals of small flares (flares 3--6 and 9) and 
large flares (1,2,7,8) to see if there was any discernible spectral difference between small 
amplitude events and larger amplitude events.  Even with these broad groupings the
number of counts in the accumulated spectra is not enough to perform a time-resolved differential
emission measure (DEM) analysis, so we adopted a different approach. 
One of the X-ray signatures of stellar flares is the creation of hot plasma, generally
{\it T}$>$10 MK, during flares.  At these high temperatures, there are relatively few atomic
transitions, mostly arising from transitions in hydrogenic and helium-like ions from high Z 
elements (Si, S, Ca, Fe).  A major contributor to the X-ray emission at such temperatures
is continuum emission due to free-free and free-bound processes.  High spectral resolution
analyses of active M star coronae reveal an underabundance in elements whose first
ionization potential is less than 10 eV relative to solar photospheric abundances \citep{vdbetal2003}.  
Abundances often increase during flares \citep{favataetal2000},
although this is not observed for all flares \citep{huenemoerder2001}.
We investigated the
spectra of quiescence, small flares, and large flares, looking for signatures
of high temperature plasma through emission lines and continuum emission,
as well as possibly elevated abundance levels during flares.  A detailed DEM analysis of 
the composite X-ray spectra, along with the STIS spectrum and other wavelength regions, is presented
in a companion paper (Osten et al., in prep.).

\begin{figure}[htbp]
\begin{center}
\plottwo{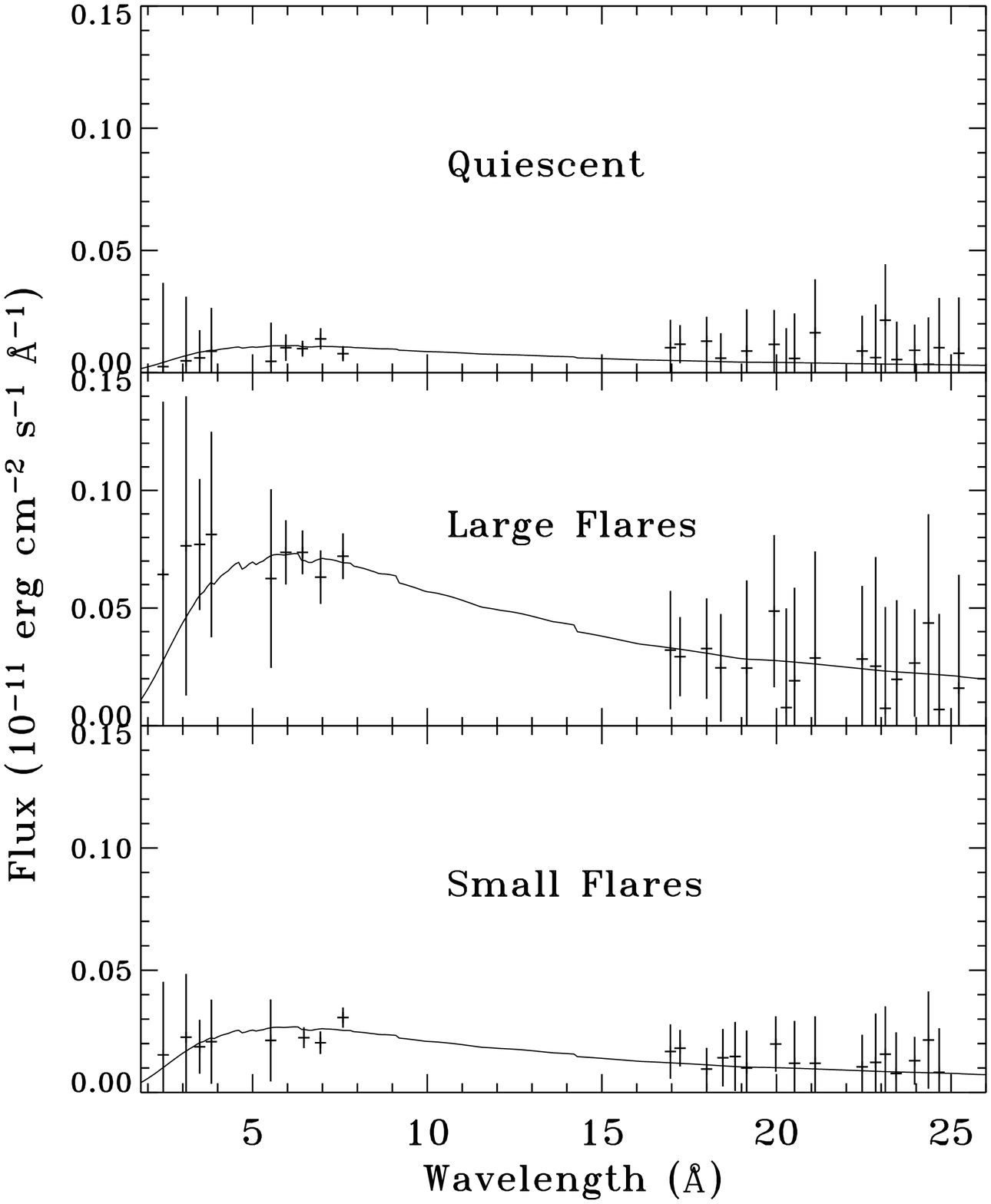}{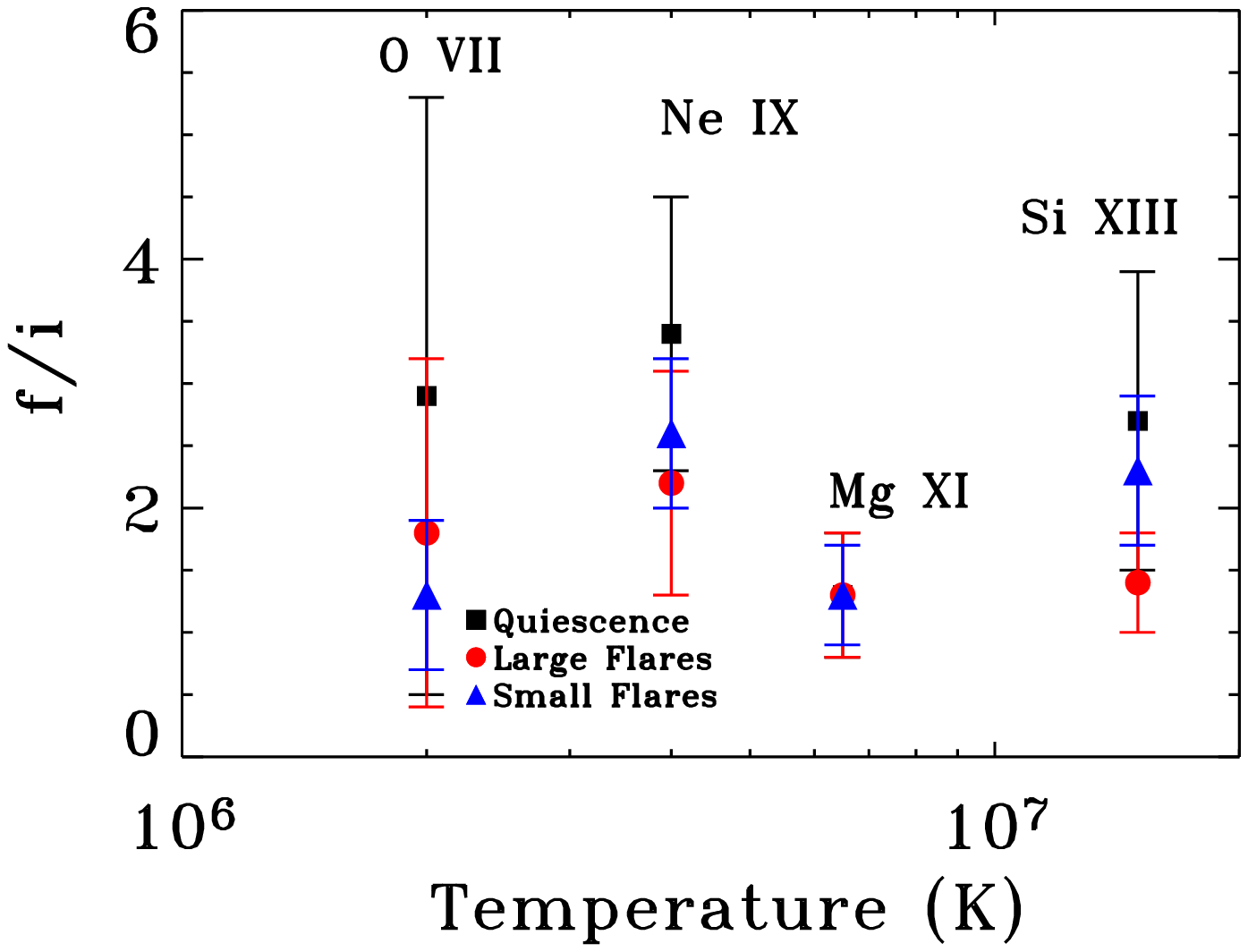}
\figcaption[]{{\bf (left)} X-ray continuum spectra during (top) quiescence, (middle) large
flares (flares 1,2,7 and 8), and (bottom) smaller flares (flares 3--6 and 9).
The intervals encompassing large enhancement flares show an increase in 
short wavelength continuum flux, associated with creation of high temperature
plasma.  The intervals encompassing small enhancement flares also shows
a slight increase compared with quiescence, signaling the existence of
hot plasma during these flares. Also shown with a solid line for all three spectra is
a schematic continuum spectrum, formed from plasma at T$\sim$12MK,
and scaled to the observed continuum fluxes at wavelengths of 10 \AA\ and longer. 
{\bf (right)} Variation in $f/i$ ratios during three time intervals:  quiescence,
large flares (flares 1, 2, 7 and 8) and small flares (3--6 and 9).  There is no evidence
for an increase in electron densities in any of the density-sensitive line ratios,
during either flare segment.
\label{chancont}}
\end{center}
\end{figure}

Figure~\ref{chancont} displays the continuum X-ray spectra during three different intervals,
corresponding to quiescence, large enhancement flares, and smaller flares.  
The wavelength regions used were determined from the Astrophysical Plasma Emission Code (APEC) linelists\citep{apecref} as comprising only weak 
emission lines (peak emissivities less than 10$^{-18}$ photons cm$^{-2}$ s$^{-1}$).
Under the assumption that most of the flux arises from continuum emission, we binned these
spectral regions to estimate the shape and amount of continuum emission.  Our previous
analyses of Chandra {\it HETGS} data \citep{ostenetal2003} have shown that this is a good approximation.
The continuum level is elevated in both flare intervals, compared to quiescence, and the 
flare continuum
enhancement is larger for the spectrum comprised of large flares.  
The continuum from large flares is about seven times
larger than quiescence; the continuum from smaller flares also shows an increase of $\sim$2.5 compared
to the average quiescent continuum shape.  
There is no discernible
difference in the shape of the continuum spectra for large flares and small flares, modulo the
large error bars at short wavelengths. 
Figure~\ref{chancont} illustrates the distribution of continuum flux from an isothermal plasma with a 
temperature of 12 MK for the three different observed activity states, where the amount of
material is varied to match the peak observed flux; 
we cannot constrain the presence of 
temperatures hotter than this amount using the continuum.
A large temperature difference would show up as a change in the peak of the continuum 
spectrum, moving towards shorter wavelengths for hotter temperatures.
The gap in continuum
from 8--17 \AA\ is due to the large number of iron L shell lines contributing emission in this region;
it is not 
possible to measure the underlying continuum emission in this region directly.

We also examined individual strong emission lines to gauge their variations
through intervals of differing activity.  Table~\ref{tbl:xrayfluxes} lists
the dominant hydrogenic and helium-like transitions of Si, Mg, Ne, and O.  The fluxes
were determined from fits to the MEG spectra after subtraction of an estimate of continuum emission (shown
in Figure~\ref{chancont}).
Gaussian line profiles of fixed width (equal to the instrumental resolution of 0.023 \AA\ for MEG) 
were used to determine the wavelength of line center as well
as total line flux; errors were estimated using the formulations of \citet{lenzayres1992}.  
Line fluxes for quiescence, large flares, and small flares are
listed in Table~\ref{tbl:xrayfluxes}.  The large flares generally also had the largest enhancements
in line fluxes.  Despite the fact that a detailed differential emission measure analysis
cannot be done on the time-resolved data, we examined evidence for temperature and abundance variations by
formulating line ratios of (1) hydrogenic and helium-like transitions from the same element,
which should have a temperature dependence but no abundance dependence; and (2) 
different elements with overlapping temperature dependences.  
These are
given in Tables~\ref{tbl:tmpratios} and ~\ref{tbl:abundratios}.  
The theoretical energy flux ratios for the corresponding
emissivities from APEC, using solar abundances of \citet{gs1998,gs1999}, are also listed.  

Table~\ref{tbl:tmpratios} lists the observed values of temperature-sensitive emission line ratios,
along with the inferred temperatures and errors derived from theoretical emissivity curves for
the transitions listed.  The four ratios indicate that a range of temperatures is present in
the coronal plasma during all three temporal segments.  For the large flares,
the ratios give evidence of hotter plasma compared to quiescent conditions; only a
modest enhancement in the temperatures is deduced during the small flares.  This may be because
the contribution from quiescent emission dominates in the small flares compared to the large flares.
The line ratios are more sensitive to small increases in the maximum temperature present, but the
continuum results are not inconsistent with the temperature changes deduced from line ratios.
This is consistent with previous analyses which indicate hotter temperatures during
larger flares \citep{feldman1995}.  The largest temperature enhancements during the flare intervals
occur for line ratios with the highest temperature sensitivity, 
which is consistent with previous observations 
showing the creation of additional
high temperature material during flares compared to quiescence \citep{ostenetal2003}.

Five different ratios which give constraints on coronal abundances
were formed, and are listed in Table~\ref{tbl:abundratios}.  The ratios were classified according
to first ionization potential (FIP) of the elements:  low FIP is less than 10 eV,
and high FIP is greater than 10 eV.  Observations of the solar corona show evidence for
a FIP-dependent abundance bias, with low FIP elements preferentially enhanced over high FIP elements
\citep{feldmanlaming2000}, while
recent observations of active stellar coronae show an inverse FIP bias \citep{vdbetal2003}.
While our observations can only constrain abundance ratios, not overall abundances, the pattern
revealed by the ratios of fluxes from low FIP to high FIP elements is consistent with an
enhancement of Ne and O (high FIP elements) over the abundances of Fe, Mg, and Si (all 
low FIP elements).  The ratios are generally consistent across the time slices, except for
a slight decrease in the Fe/Ne and Mg/Ne ratios during the two flares, which is consistent with
a slight enhancement of the Ne abundance.  During the time interval encompassing 
small flares, there is a noticeable 
increase in the Si/Mg abundance ratio compared both to quiescence and larger flares.

We also examined evidence for variation in electron densities during the spectra comprised
of large and small flares, respectively, from quiescent conditions.  
The intercombination lines $i$ and
forbidden lines $f$ of helium-like triplets of
O~VII, Ne~IX, Mg~XI, and Si~XIII can be used to estimate the electron density
at temperatures ranging from 2 to 15 MK.  The energy flux ratios $f/i$ are plotted
in Figure~\ref{chancont} at the temperature of peak emissivity
for the ion; the flare intervals show no significant variation 
from the quiescent points.  

\section{Multi-Wavelength Correlations}
There are two temporal regions of overlap between the five telescopes involved in
the campaign on 20 September 2001: VLA, McDonald 2.1, 2.7m telescopes, and {\it Chandra} 
overlap from 02:00--12:00 UT,
and {\it Chandra} and {\it HST} overlap from 16:00--24:00 UT.

\begin{figure}[htbp]
\begin{center}
\scalebox{1.0}{\plottwo{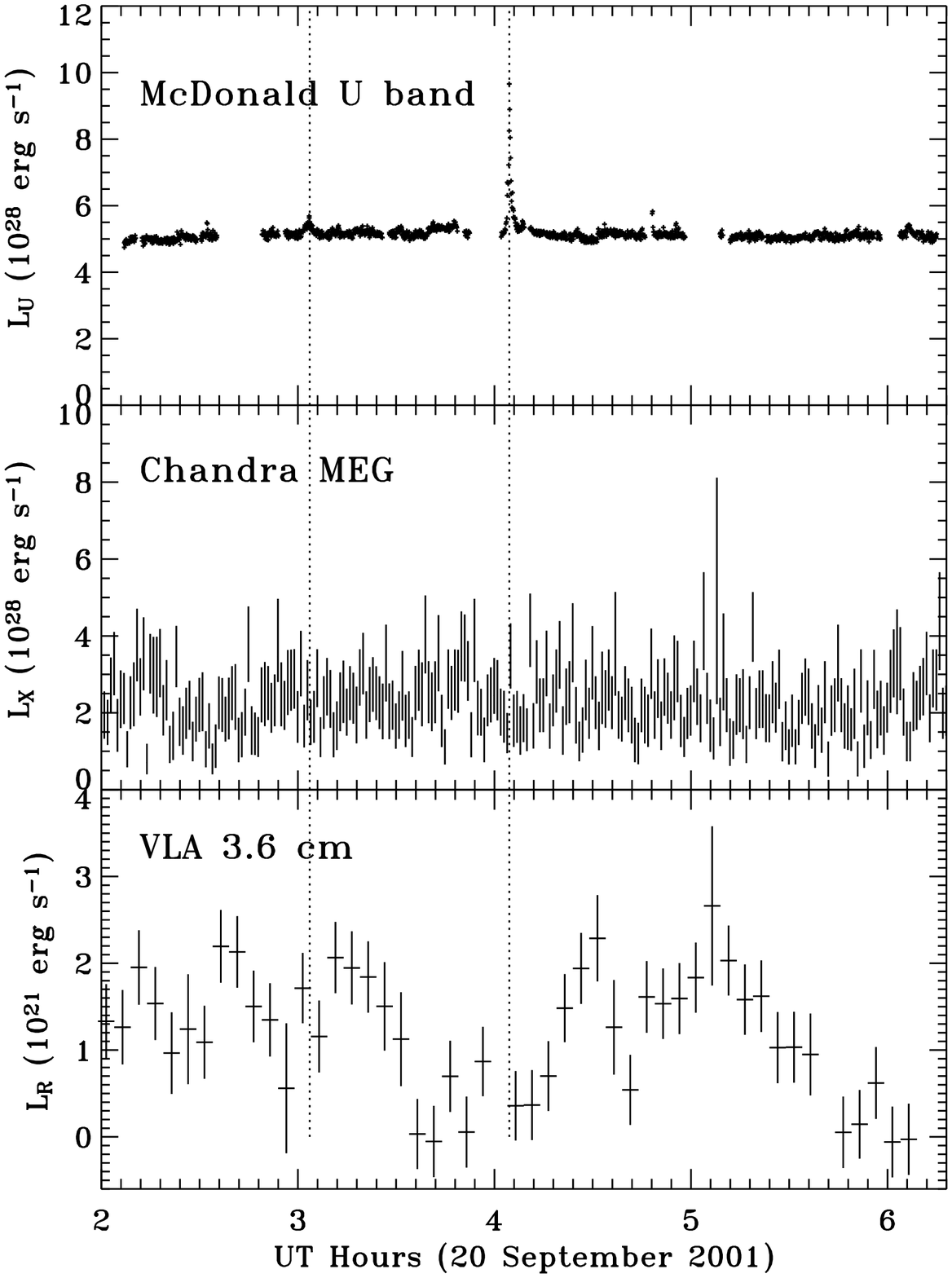}{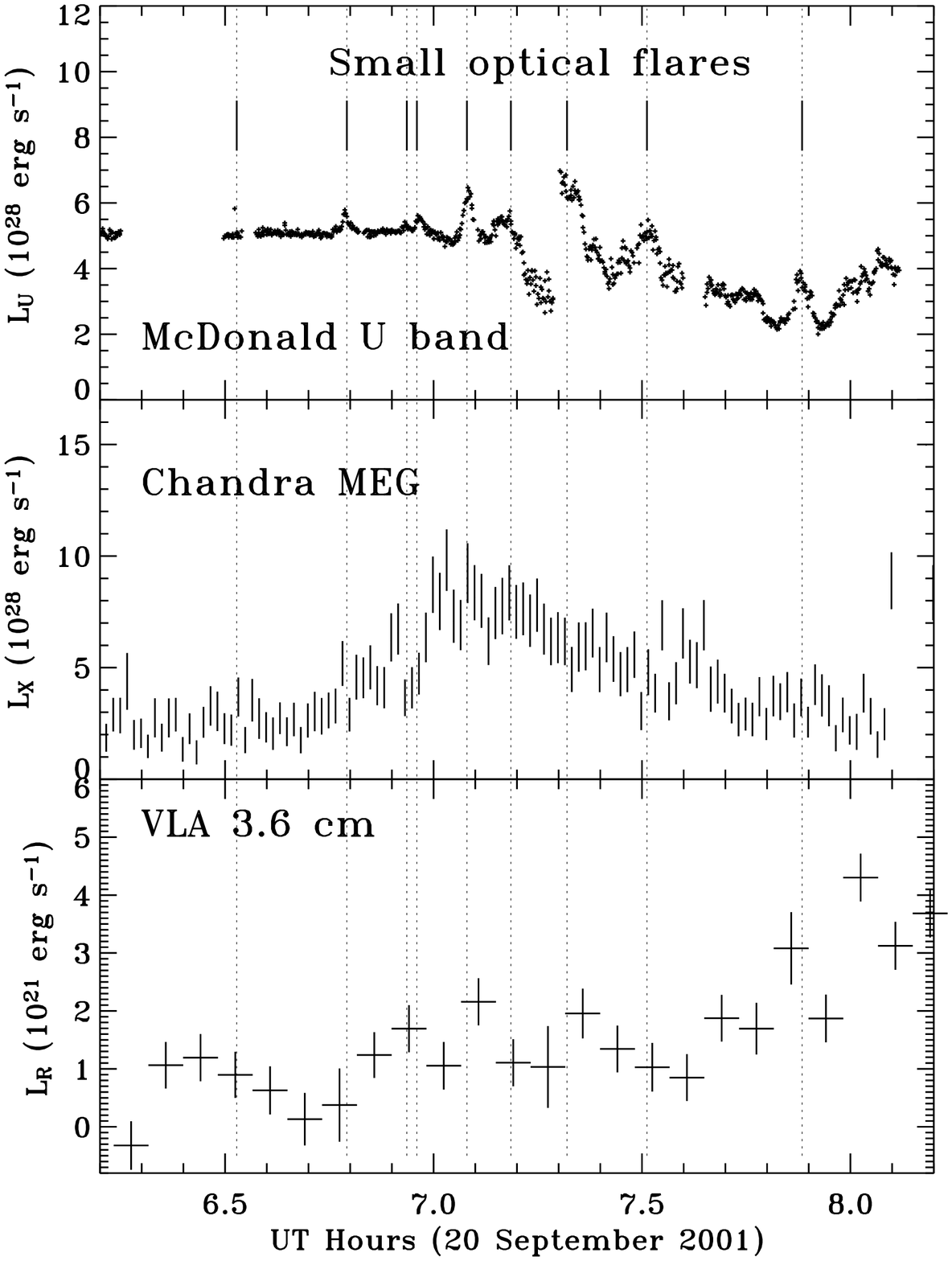}}
\figcaption[]{Light curves of optical, radio, and X-ray data during time of overlap.  
Radio binning is 300 seconds; X-ray binning is 60 seconds; optical binning is 2.839 seconds, with a cadence
of 7 seconds for multiple filter monitoring and filter wheel rotation.  Dotted lines indicate
times of peak optical emission.
 \label{multiw1}}
\end{center}
\end{figure}

Figure~\ref{multiw1} illustrates the behavior from 02:00 until 08:00 
during the
first period of overlap.   From 02:00--06:00 there 
is a small event around 03:04, and a moderate optical flare at 04:04. 
The X-ray variations occur with an average 
and standard deviation of 0.17$\pm$0.03 counts s$^{-1}$, which we identify as the ``quiescent''
count rate.  There are a few small radio (3.6 cm) enhancements, which do not appear to be correlated
temporally with any of the optical events.  As mentioned in \S 3.1, 
the 6 cm variations from roughly 00:30 until 04:00 are characterized by large amounts of
circular polarization, yet there is no multi-wavelength context for these variations as
no corresponding variability is evident at X-ray or optical wavelengths.  
This is generally
consistent with previous determinations of a lack of correlation between coherent flare signatures
and X-ray/optical flare signatures \citep{kunduetal1988,gagneetal1998}.

\begin{figure}[htbp]
\begin{center}
\scalebox{0.5}{\plotone{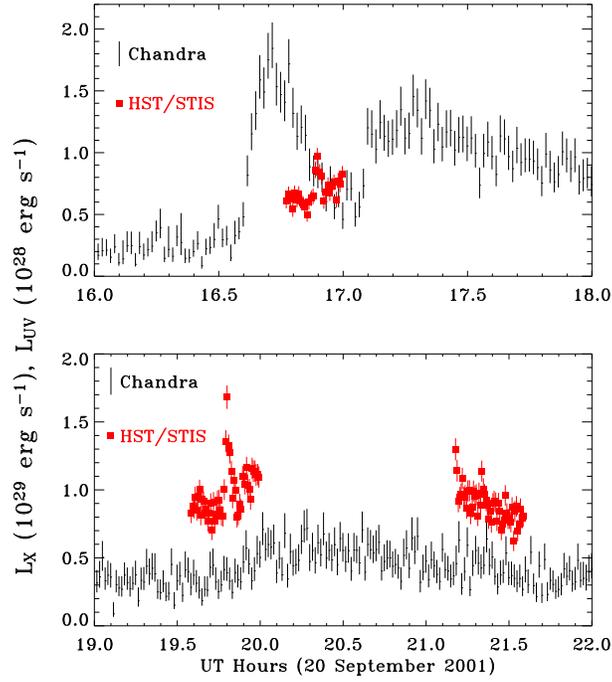}}
\figcaption[]{Close-up of two small UV enhancements during a time of X-ray
flaring activity.  Top panel shows X-ray flares 7 and 8; bottom panel, X-ray flare 9.
Binning in both light
curves is 60 seconds. Lines indicate $\pm$1$\sigma$ uncertainties in 
luminosity.  Neither the time scales nor the energetics of the X-ray/UV events is consistent,
and appears to be a random association of flares; see text for details.
\label{chanhst}}
\end{center}
\end{figure}

Figure~\ref{multiw2} shows the behavior from 08:00--12:00 UT during the first period of overlap.
The first impulsive X-ray flare 
occurred during an interruption in optical coverage, and there are
no apparent radio variations during this event.  
There are subsequent small X-ray
enhancements, but any 
possible corresponding variations in the optical
are obscured by heavy cloud coverage, and in the radio by the decay of a large event.
Figure~\ref{chanhst} illustrates the X-ray and UV variations during the time of overlap of 
{\it Chandra} and {\it HST}.
The two small UV flares appear to occur during periods of flaring X-ray activity.

\subsection{Large Radio/Optical Flare}
\begin{figure}[htbp]
\begin{center}
\scalebox{1.0}{\plottwo{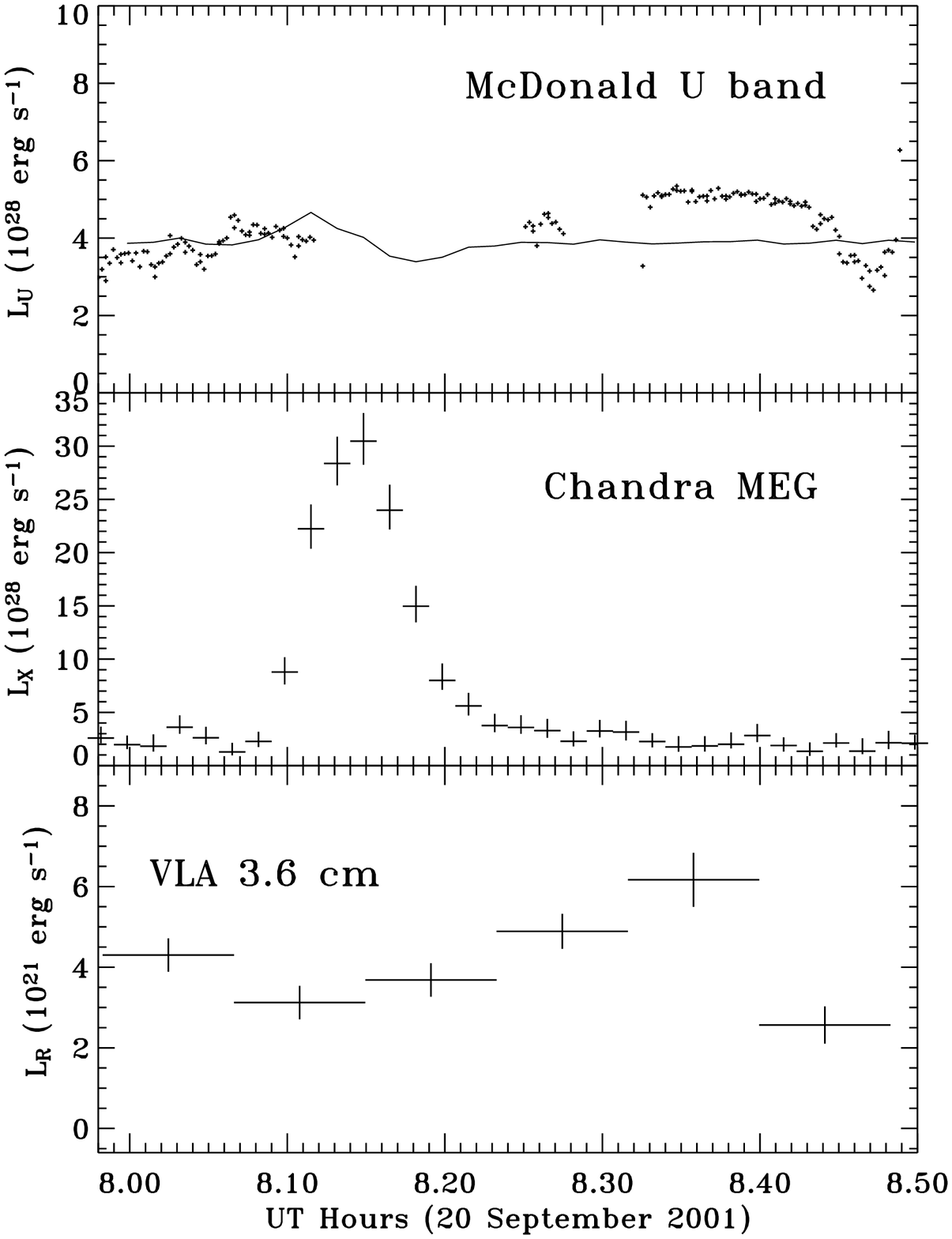}{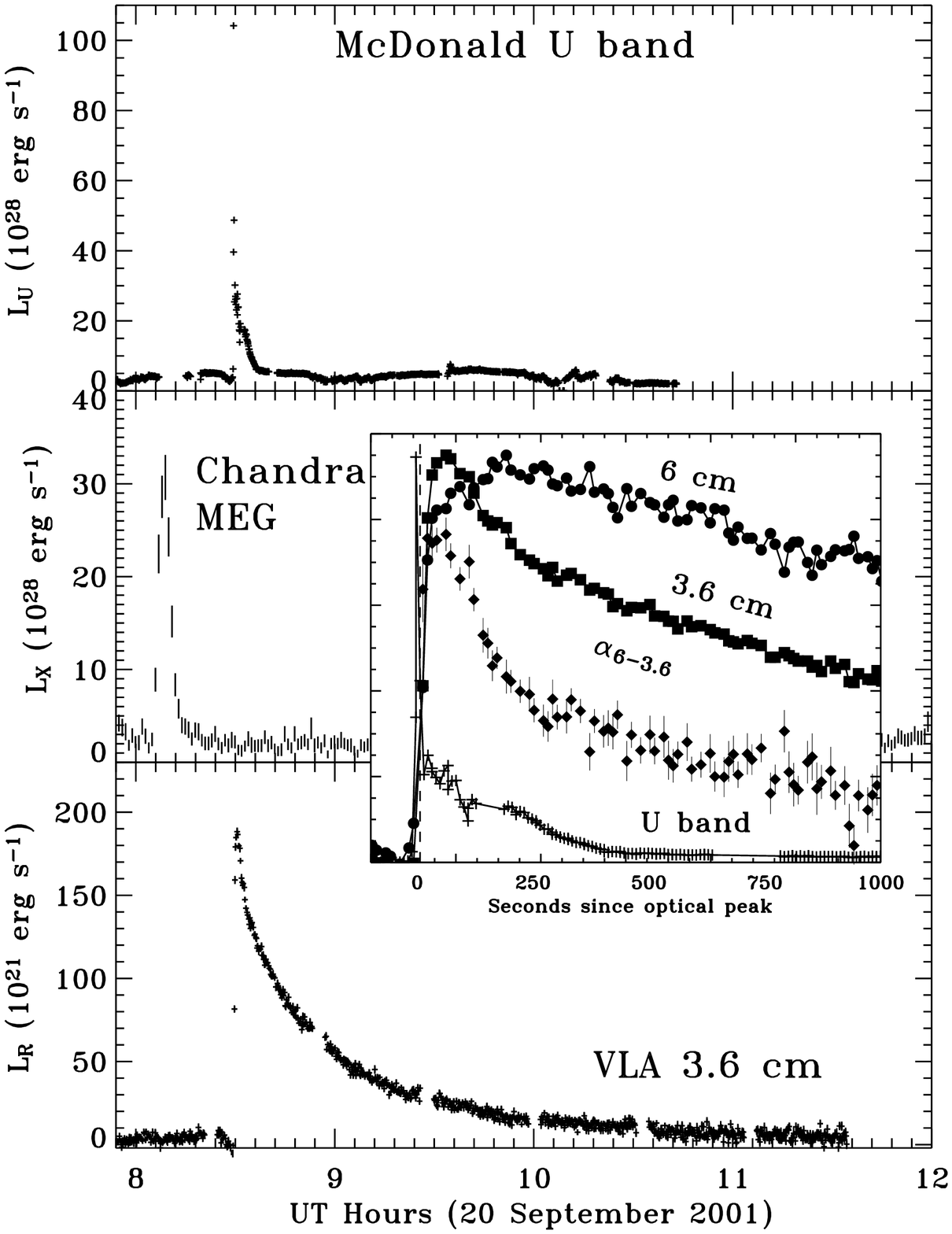}}
\figcaption[]{Light curves of optical, radio, and X-ray data during time of overlap from
08:00--12:00 UT.
X-ray binning is 60 seconds; optical binning is 2.839 seconds
with 7 second cadence. In the left panel, radio binning is 300 seconds; in the
right panel, it is 10 seconds. The upper left panel overplots the time derivative of
the X-ray light curve, to compare against optical variations in light of the Neupert
effect.  The inset to the right panel details the evolution of an impulsive
flare seen at radio and optical wavelengths in normalized flux units. 
The dashed line indicates the location in the
U band light curve where the first derivative changes sign after the flare peak; this may
signal a change from the impulsive to gradual phase of the flare.  Note that according to this 
criterion, the radio flares occur during the gradual phase of the flare.  The diamonds indicate
the spectral index between 6 and 3.6 cm during the initial stages of the decay.
See section 4 for more details.  
\label{multiw2}}
\end{center}
\end{figure}

There is a close temporal association between the large
U band flare and the radio flare displayed in Figure~\ref{multiw2}, yet no indication of an accompanying X-ray flare
exists.   The optical flare peaks $\approx$ 20 minutes after the impulsive X-ray event,
too far apart in time to causally connect under any realistic length scales or velocities.
The inset of Figure~\ref{multiw2} shows a close-up of the U band and radio
flares; all three flares appear to start at the same time, within the 10 second temporal
sampling of the radio data. 
The U band peak precedes the 3.6 cm peak by roughly 54 seconds.
\citet{hawleyetal2003} used as a criterion
to determine the changeover from impulsive to gradual phase of the flare 
the time when the first derivative of the U band light curves changes sign following 
the flare peak. 
By this definition, the 
impulsive phase ends approximately 7 seconds after the U band peak and most of the 
radio flare at 3.6 and 6 cm occurs during the gradual phase of the flare.  

Solar white light flares are rare, but seem to show evidence of association with
hard X-ray and microwave bursts, suggesting that the optical continuum emission is at least partly
due to the acceleration of electrons \citep{neidig1989}.
Our observations of a temporal association between an energetic optical stellar flare and a large
microwave flare seem to support this hypothesis for the outer atmosphere of EV~Lac.
\citet{neidigkane1993} observed similar timescales for
the hard X-ray (50 keV) and white light solar flare emissions, with the hard X-ray
emission generally occurring first, by $\approx$ 8 seconds. 
If we assume that the optical continuum emission is a response to 
the bombardment of EV~Lac's lower atmosphere by accelerated particles, and the microwave
emission is gyrosynchrotron emission from accelerated electrons caught in a magnetic trap, 
then the time delay of the 
gyrosynchrotron emission could be due to the kinematics of the particle injection and trapping.
An anisotropic pitch angle distribution of the injected electrons can cause a significant 
delay between the time profile of injected electrons and that of trapped electrons, as \citet{leegary2000}
demonstrated for solar flares.  Without more constraints, we cannot say conclusively that this
is the case for the flare on EV~Lac, but the evidence is suggestive.  

While we do not have conclusive 
multi-band optical photometry during the flare due to cloudy conditions, we assume that the
characteristics of this flare are similar to other well-studied flares on similar stars
\citep{hawleyetal2003};
namely, that the rise in optical continuum emission is due to black-body radiation at 
temperatures of about 10$^{4}$K, covering a small area fraction ($\sim$few $\times$10$^{-4}$)
of the stellar disk.  The temporal association between the optical and radio flares suggests that
the two are causally related.  The radio flux density measures the brightness temperature and 
source size, both of which are unknown. 
 We use the optical area filling factor derived from other flare studies as appropriate
for this flare, and assume that the source size expands from the photosphere (where the optical emission
originates) to the corona (where the radio emission originates).  
We assume that the photospheric magnetic field involved in the flare is
of the same order as deduced from observations of \citet{jkv1996}, namely, 2.5--3.5 kG, and constrain
the coronal magnetic field by the harmonic numbers appropriate for gyrosynchrotron emission
($\nu=s \nu_{B}$, $\nu_{B}=$2.8$\times$10$^{6}$B Hz, $s=10-100$, and $\nu$=4.9, 8.4 GHz). 
Then we can derive an approximate
relation for the size of the radio source, using potential field configuration and 
conservation of magnetic flux.  The area expansion between photosphere and corona should then be \\
\begin{equation} \label{eqn:gamma}
\Gamma=\frac{B_{phot}^{1/2}}{B_{cor}^{1/2}} = \frac{A_{cor}^{1/2}}{A_{phot}^{1/2}} 
\end{equation}
where $B_{\rm phot}$ is the photospheric magnetic field strength, $B_{\rm cor}$ is the 
coronal magnetic field strength, $A_{\rm cor}$ is the cross-sectional area which encompasses
the coronal magnetic field, and $A_{\rm phot}$ is the cross-sectional area which
encompasses the photospheric magnetic field.  Using the above estimates of $B_{\rm phot}$
and $B_{\rm cor}$ gives an estimate of $\Gamma$ ranging from 4--10.
If {\it A$=${\it x}$\times$A$_{\star}$}, 
and $x_{phot}$ is $\sim$ 10$^{-4}$, then
$x_{cor}$ should be 10$^{-3}$--10$^{-2}$.  
\citep[We note in passing that results from solar
studies have indicated much smaller area expansion factors, $\Gamma$ of order 1.3;][]{klimchuk2000,watko2000}.
For the apparently optically thick conditions at the flare start, 
the brightness temperature can be equated with the effective temperature of the electrons;
we use the 6 cm fluxes as they appear to be optically thick during all of the flare decay.
The implied brightness temperatures are $\sim$ 10$^{12}$K, 
and the average energy
of the accelerated electrons is quite high, $\sim$ 90 MeV.  
However, VLBA studies of dMe stellar coronae
show \citep{benz1998} a source size several times the stellar radius, and {\it T$_{b}$} of a 
few $\times$10$^{8}$K, with loop top field strengths $>$ 15 G.  
In order to obtain a brightness temperature within the expected range for 
gyrosynchrotron emission at centimeter wavelengths, {\it T$_{b}$} $\approx$ 10$^{8}$--10$^{10}$K,
the area filling factor for the radio source must be of order unity; i.e. the radio source must be
the size of the star itself.  
The average accelerated
electron energy would then be in the range several tens -- several hundreds of keV.
This is consistent with the results from \S3.1.1.

The estimate of the energy
of the accelerated particles becomes important in quantifying the ambient density of the magnetic
trap. Following equation~\ref{eqn:taud}, electrons with average energy $\sim$ 90 MeV imply high
ambient electron densities (n$_{e} \sim$ 10$^{12}$ cm$^{-3}$) 
to explain the observed decay timescales as trapping effects.  These large
densities would imply substantial amounts of free-free absorption, and make such an
interpretation problematic.
For brightness temperatures 
compatible with the large observed VLBA source sizes, and energies $\sim$ few hundred keV, a much lower 
ambient electron density is required, n$_{e}\approx$ 10$^{8}$ cm$^{-3}$. 
Without an actual measurement of the radio source size, both solutions are possible,
but the latter seems more plausible.

\subsection{The Breakdown of the Neupert Effect}
The coordination of so many telescopes spanning different wavebands offers a unique
opportunity to test current understanding of stellar flares.  
Within the standard framework of solar flares \citep{denniszarro1993}, the optical and radio emission 
probe the energetics of accelerated electrons.  Solar (and stellar) ultraviolet emission 
enhancements are associated with nonthermal energy deposition, and an increase in
thermal coronal radiation
occurs as result of chromospheric evaporation, in which chromospheric plasma,
having been heated to coronal temperatures by the deposition of nonthermal energy from
accelerated particles, undergoes a radiative instability and fills up coronal loops
to radiate at EUV/X-ray wavelengths.  This gives rise to an observed
relationship between thermal coronal emission and radio/hard X-ray nonthermal emission, with the
timescale for thermal coronal radiation 
being proportional to that for the increase in nonthermal energy deposition.  
Despite the success with which previous authors
found evidence of the Neupert effect in dMe flare star coronae 
\citep{hawleyetal1995,gudeletal1996,gudeletal2002,hawleyetal2003}, 
we find a complete breakdown between the action of nonthermal particles
and thermal response, on numerous occasions.

\begin{figure}[htbp]
\begin{center}
\scalebox{0.5}{\plotone{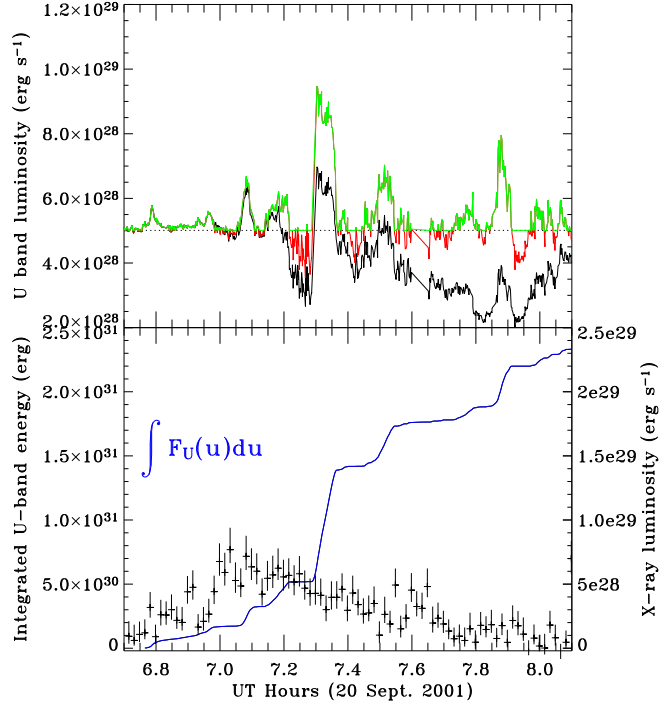}}
\figcaption[]{{\bf (top)} Optical U band light curve, converted to 
luminosity (using quiescent luminosity of 5.01 $\times$10$^{28}$erg s$^{-1}$,
shown as dotted line).
Black curve shows original data; 
red curve sketches data after cloud
correction.  The green curve sets any points falling 
below the quiescent luminosity equal to it; the red and green curves are equivalent
above the quiescent luminosity. 
{\bf (bottom)} Crosses show the X-ray luminosity variations during the time
of a gradual flare (X-ray flare 1 in Figure~\ref{chanlc}; quiescent luminosity 
has been subtracted, time binning is 300 seconds) and blue curve
delineates the time integral of U band flare luminosity 
(quiescent luminosity subtracted). The profile of the integrated U band 
luminosity does not match the instantaneous X-ray luminosity, as one would expect
under the Neupert effect (see text). \label{neupert}}
\end{center}
\end{figure}

A gradual X-ray flare is evident during the interval 06:00--08:00 in Figure~\ref{multiw1}, along with
several small optical flares. There is no noticeable response at radio wavelengths to
either the optical or X-ray variations.  We investigated the luminosity variations in the 
optical light curve to determine if their time variations were consistent with
a Neupert effect interpretation.  
The models of \citet{hf1992} showed that 
the optical stellar flare continuum emission can serve as a proxy for 
hard X-ray nonthermal emission, which arises due to the particle
acceleration event at the beginning of the flare.  
\citet{hawleyetal1995} and \citet{hawleyetal2003}
found a correspondence between
the time integral of the U-band luminosity and the instantaneous EUV luminosity, for
flares on another active flare star (AD~Leo).  
Figure~\ref{neupert} illustrates the X-ray and optical data during the time of the first X-ray flare.
Because of non-photometric conditions, we attempted to correct for the
influence of clouds by setting the luminosity during apparent quiet times
equal to the U-band luminosity of \citet{gc1969}, 5.01$\times$10$^{28}$ erg s$^{-1}$.
During intervals affected by clouds (e.g. 7:06--8:12 UT), 
we fit the temporal variations with a second order 
polynomial and divided the observed variations by this fit, normalizing to the 
quiescent U-band luminosity.  Any subsequent variations below this value were reset
to the quiescent luminosity.  The top panel of Figure~\ref{neupert} illustrates this
sequence.  The bottom panel shows the time integral of the corrected U band luminosity,
overlaid on the X-ray luminosity variations.  
Both the X-ray light curve and integrated U-band
luminosity have had estimates of the quiescent luminosity subtracted.  The integrated U-band 
luminosity does show a gradual increase during the corresponding gradual increase of the 
X-ray flare, yet the peak of the X-ray variations occurs prior to a small plateau in the
optical energy.  There is a complete breakdown in the apparent relation between the two after
this, however;
the presence of optical enhancements after the peak of the X-ray flare accounts for the
increase in the integrated optical luminosity.  It appears that the Neupert effect does not
provide a wholly satisfying framework on which to interpret such variations.

For the impulsive X-ray flare shown in Figure~\ref{multiw2}
the Neupert effect can be investigated
by comparing the time derivative of the X-ray variations with the instantaneous optical/radio
light curve.
We compared the time derivative of the X-ray
light curve with the optical variations; there is no correspondence and thus it appears that 
the U band data do not track the start of the 
X-ray flare.

As spectacular as the optical/radio flare depicted in the right panel of Figure~\ref{multiw2} is, 
it is even more remarkable that no 
X-ray flare seems to occur within the tens of minutes on either side of the time of peak flux enhancement.
There is inconclusive evidence of extended
chromospheric emission during the flare, and a definite lack of thermal coronal flare emission.
One possible explanation for the lack of an X-ray signature follows from
the long observed timescales for radiative decay of the radio emission, which implies low 
ambient electron densities (for
particle energies of a few hundred keV, n$_{e}$ would be $\sim$ 10$^{8}$ cm$^{-3}$).
The n$_{e}^{2}$ dependence of the thermal X-ray emission
preferentially biases high density structures;  
quiescent X-ray structures in active M dwarfs have 
n$_{e} \geq$ 10$^{10}$ cm$^{-3}$ 
at temperatures T$_{e} \geq$ 3MK \citep{vdbetal2003}
and would explain why X-ray flare emission from lower density structures wasn't seen.  
Another possibility invokes heating at
lower temperatures than probed by Chandra: a stellar analogue of solar transition 
region explosive events \citep{dereetal1989}.
\citet{ayres2001} interpreted an enhancement in transition region emission and observed
lack of X-ray (and radio) signature
on the active binary system HR~1099 as a possible stellar analogue of such an event.
Another alternative could be excellent trapping efficiency of the radio loop, 
so that very few particles
precipitate from the trap to heat the surrounding chromosphere and produce subsequent evaporation.
An exotic energy budget could also be invoked, where most of the energy goes into accelerating particles (perhaps
protons) which would allow deeper penetration of the photosphere and produce continuum emission, but
no line emission nor significant evaporation.  
While we have no conclusive answer to the puzzling
flare examined here, it is an example of a unique kind of stellar flare which needs to be
considered in the zoo of stellar flare phenomena. 

The temporal association between the two UV flares and two corresponding X-ray flares 
depicted in Figure~\ref{chanhst} invites the
interpretation that they are causally related. 
Flare-related enhancement in UV line emission can arise
from a deposition of nonthermal energy, as both observations of solar flares \citep{cheng1999}
and theory of stellar flares \citep{hf1992} suggest.  
If this is the case, then the UV flares are a proxy 
for nonthermal energy deposition.
We could not successfully use the 
Neupert effect interpretation to explain the temporal relationship between the
UV/X-ray flares seen here, however,
as there is a large mismatch between the timescales over which the UV flares operate (4-5 minutes)
and the timescales for the decay of the X-ray radiation (2--3 hours).
The UV/X-ray emissions could still be manifestations of the same
reconnection event, but not with a simple temporal association as the Neupert effect prescribes.

\subsection{Flare Statistics: Causal Connections based on Temporal Associations}
The astounding lack of correlation between flares seen in different wavelength
regions in this multi-wavelength flare campaign is puzzling, and we investigated
the idea that the temporal associations are based solely on chance.  If the 
distribution of flares is random in time, then 
the time between flares (waiting time distribution)
should also be a random process, and described by a probability density distribution of the
form P($\Delta$t)=$\lambda e^{-\lambda \Delta t}$, where $\Delta t$ is the time
between flares and $\lambda$ is the mean flaring rate.  
The waiting time distribution has been studied for the
case of solar flares and stellar optical flares, and is consistent
with a Poisson distribution for small waiting times \citep{wheatland2000,lme1976}.   
The probability of observing N flares given a time interval $\Delta t$, with mean flaring rate 
$\lambda$ is then $P(N;\Delta t, \lambda)=e^{-\lambda \Delta t} (\lambda \Delta t)^{N}/N!$.
If flaring in two disparate wavelength regions occurs independently, then we can
multiply the individual probabilities together to estimate the likelihood of seeing
such a temporal association by chance.


The X-ray flaring rate determined from the Chandra data described here (0.32 flares hr$^{-1}$) 
is consistent with U band flaring rate range of 0.10--0.40 flares hr$^{-1}$ 
found by \citet{letoetal1997} for EV~Lac.
There is an apparent disconnect with the optical flaring rate estimated during
our campaign: $\sim$11 optical flares in 8.6 hours, or 1.3 flares hr$^{-1}$ (clouds prevent
an accurate count).  The flaring rate in the UV is less certain due to the small 
time coverage (2 flares in $\sim$3 hrs), but we estimate it as 0.66 flares hr$^{-1}$.
The radio flaring rate is likewise uncertain, but appears to be greater than 0.09 flares hr$^{-1}$,
due to the one large flare and possibly many small flares observed in $\approx$ 11 hours.
In the 1.5 hour period between 06:30 and 08:00 UT shown in the right panel of Figure~\ref{multiw1},
there are approximately 9 optical flares and 1 X-ray flare; the probability of a random
occurrence of the nine optical flares and one X-ray flare is 4.7$\times$10$^{-5}$, 
and yet the Neupert effect in its simplest
formulation cannot be used to explain the pattern of energy release.  
For the $\sim$ 0.5 hr interval between each UV/X-ray flare pair, the probability
of having a random flare association between a UV flare and an X-ray flare using the above estimates for the flaring rates
is 0.032.  As shown in Figure~\ref{chanhst}, in the six hours from 16:00-24:00 there are 3 X-ray flares and 2 UV flares;
the probability of all these flares being randomly associated is 0.026.  
Thus we cannot reject the hypothesis that the $<$ half-hour interval between the two
UV/X-ray flares is due purely to chance, based on the high flaring rate and the low
number of UV/X-ray flare associations.
Both the optical flare and the radio flare depicted in Figure~\ref{multiw2}
were large events, and the statistics of
the frequency distribution of flare energies in the optical \citep{letoetal1997}
suggest a frequency of 0.022 flares hr$^{-1}$, given
an estimated U-band energy of E$_{U}$=5.2$\times$10$^{31}$ erg.
The probability of observing an optical flare of this energy and a radio flare, using the estimated
radio flare rate from this observation, within 60 s is 5.5$\times$10$^{-7}$.
Thus it appears that the radio and optical flares are not randomly associated, but
the lack of a corresponding X-ray flare signature defies our current understanding
of stellar flares.

These estimates assume that all flares occur with the same frequency; 
studies of optical flares reveal that the flare frequency depends on flare energy \citep{lme1976},
and investigations of EUV/SXR variations of other M dwarf flare stars \citep{gudel2003} also shows 
a distribution of flare frequencies with flare energy.  Therefore, a more detailed
treatment of expected and observed multi-wavelength flare associations 
must be performed before one can
assign any significance to one (or two) pairs of flares occurring within a reasonably 
small amount of time.  It is especially important to develop a causal model relating the
energies or peak luminosities of flares in different wavelength regions, in order to 
constrain random associations.

\section{Discussion}
The study of flare stars continues to reveal new surprises, and our
campaign on EV~Lac was no exception.
The multi-wavelength observations detailed in this paper show both
frequent and extreme levels of variability in the dMe flare star EV~Lac. 
At radio wavelengths, we observe highly polarized emission at 6 cm,
which we conclude is the signature of coherent emission processes occurring
in EV~Lac, and supports previous longer wavelength observations showing a large 
circular polarization signature.  We also observed an intense radio outburst
which was evident at two frequencies; the timescales and polarization signatures
suggest that this is an extreme example of a flare emitting gyrosynchrotron 
radiation from accelerated particles.  The temporal behavior of the spectral 
index shows large, steep values during the flare rise and peak, suggestive of
optically thick conditions.  The spectral index is constant for about 40 seconds 
during the time of radio flare rise; such a signature may be indicative of injection
of accelerated electrons into a magnetic trap.  The decay of the radio radiation
on long timescales is consistent with the trapping of accelerated particles and
subsequent energy losses due to Coulomb collisions.

\citet{leegary2000} demonstrated based on solar flares that a constant spectral index can be
obtained from the beginning of particle injection into the flare loop until the time 
when the maximum number of particles is injected.
\citet{leegary2000} also showed that an initially beamed pitch-angle distribution can result in
an offset of the time of maximum of gyrosynchrotron radiation relative 
to the time of maximum injection; this is due to the extra time needed to fill all
the pitch angles required for the resonance condition.
This may explain the
optical/radio correlations found in our campaign, if the optical emission is a signature
of the particle injection, the flattening of the radio spectral index during the 3.6 cm
flare rise is related to the injection profile, and the temporal offset of the optical
flare maximum from the gyrosynchrotron maximum is a result of beaming in the
pitch-angle distribution.  

Intense centimetric variations in other M dwarf flare stars 
are usually accompanied by large values of circular polarization,
are shorter lived,
and are
attributed to coherent mechanisms;
for example, DO~Cep (dM4e) was observed by \citet{whiteetal1989} in
a radio flare at 6 cm which attained a maximum flux density of 80 mJy
(L$_{R}\sim 1.5\times10^{15}$ erg s$^{-1}$ Hz$^{-1}$), lasting for $>$ 450 seconds.
This flare, however, was almost 100\% circularly polarized, and was narrow banded, occurring
most strongly at 4535 MHz but hardly at 4985 MHz; the star was undetected during a second 
epoch observation.
No emission mechanism was suggested.
\citet{stepanov2001} reported a large
amount of circular polarization and flux density (peaking at $\approx$ 300 mJy, or L$_R\sim
8.7\times10^{15}$ erg s$^{-1}$ Hz$^{-1}$)
on AD~Leo (dM2e) at 6 cm, which they interpreted as plasma emission; this burst, however, had a
very short duration ($\sim$ 1 minute).
And \citet{benz1998} saw large amounts of right circular polarization on UV~Cet (dM5.5e)
at 3.6 cm (maximum Stokes V flux of $\sim$ 30 mJy [L$_{R}\sim 2.4\times10^{14}$ erg s$^{-1}$ Hz$^{-1}$], total duration $\leq$ 15 minutes)
which \citet{bingham2001} attributed to emission from a cyclotron maser instability.

There is a signature of highly polarized emission from EV~Lac, occurring at 6 cm, with timescales
of hours, and peak radio luminosity L$_{R} \sim$3$\times$1$^{13}$ erg s$^{-1}$ Hz$^{-1}$.
This behavior is more similar to the types of enhancements seen in other dMe flare stars
which are attributed to a coherent mechanism, 
yet the variations described here are of smaller luminosity and occur over longer
timescales of hours.  
\citet{whiteetal1989} reported large amounts of circular polarization ($>$80\%) at 
20 cm wavelengths; however EV~Lac was undetected in their 6 cm measurements, and there are no
other reports of measured circular polarization from this star.  The 
enhancements with large circular polarization signature could be connected to the intense
but brief events reported on other active dMe stars.

Several small flares are seen in the optical and ultraviolet.  
Two small UV enhancements were found during the HST observations.  Both had approximately
the same enhancement above the surrounding count rate in integrated UV light (and roughly 
similar durations), 
and we were able to determine the characteristics of the two flares from line profile variations
of bright transition region emission lines.
There was 
no evidence for bulk plasma motions greater than a few km s$^{-1}$,
and the amount of turbulent broadening was different in the two flares.

The X-ray data show a panoply of flare types, with flare timescales ranging from
a few minutes to several hours.  Analysis of undispersed spectra corresponding to
individual flares show no evidence of abundance increases, and only moderate temperature
enhancements.  For three flares which lasted sufficiently long, we separated the rise phase
from the decay phase and found a flare with no temperature increase during
the rise, one with a modest temperature increase (7 MK),
and a third with a dramatic temperature increase ($\geq$ 30 MK).
A comparison of temperature-sensitive
emission line ratios in the X-ray grating spectra reveal evidence for higher temperature
plasma during flare intervals, with the largest temperatures occurring during the
time intervals with the largest enhancement events.  We are able to isolate
the shape and level of the continuum emission, and find that it agrees qualitatively
with the information derived from the temperature-sensitive line ratios.
Abundance-sensitive line ratios 
cannot constrain the coronal abundance directly, but do suggest a pattern reminiscent
of an inverse-FIP effect, as is (now) commonly seen in the coronae of active stars.
There is no evidence for large changes in the abundance ratios during times of flares.

In the multi-wavelength flare associations, we find a complete breakdown of 
expected relations between the signature from accelerated electrons
(radio, but secondarily optical and ultraviolet) and the signature presumably
due to plasma heating from the accelerated electrons (soft X-ray emission).
We find an optical flare coincident with a radio flare to within 54 seconds, 
which cements the
association of optical stellar flares with a bombardment of the lower atmosphere by
nonthermal electrons.  
Particularly puzzling is the lack of an observable X-ray enhancement with this radio/
optical flare, which would have signaled the influence of the nonthermal particles in heating
chromospheric material to coronal temperatures via a Neupert effect
relationship.  
Lacking this signature, we can only speculate about the dynamics/characteristics/energy
budget of this unusual flare: perhaps the structure involved in the radio/optical flare
was of too low an ambient electron density to be an efficient radiator of soft X-rays, and
thus was invisible to us.

There are other temporal intervals where optical and UV enhancements could not 
be connected to contemporaneous X-ray flaring through either timescales or
energetics.  In only one case can we conclude that a chance superposition 
of UV and X-ray flares was occurring; in the other, we have no way to explain 
the correlation using simplistic models.
We suggest that a more sophisticated, statistical treatment of multi-wavelength
flare correlations is necessary to make causal connections based on temporal associations,
particularly for stars (like EV~Lac) with high flaring rates.

\section{Future Work}
The multi-wavelength flare associations described in this paper point to a complex
view of the stellar flare process.  Flares are found in every wavelength region in
which we observed, and yet multi-wavelength correspondences between flares
are elusive for all except the largest optical and radio flares.
The state of understanding of stellar flares has been riddled with
inconsistencies between the solar ``paradigm'' and stellar reality, which often reveals a
confounding set of multi-wavelength interrelationships.  
Yet the Sun itself is extremely complex, and general trends have
only been gleaned after decades of intense study.  The stellar community must therefore assess the
importance of any one result in light of the spotty nature of our observing campaigns.

The radio flare presented in this paper has been intriguing, though frustrating, to study,
partly due to the choice of radio frequencies and resulting optical depth effects present
in the data.  This points to the need for 
radio observations at higher frequencies to establish injection and trapping properties
of stellar coronal loops, at frequencies where the emission is predominantly optically thin.
With the advent of the Expanded Very Large Array (EVLA), which will have a factor of $\sim$10
increase in sensitivity over the current VLA,
it will become feasible to perform high time-resolution multi-frequency observations such as those
described here, at higher, optically thin, frequencies (15 -- 40 GHz) and therefore to probe the dynamics of
flares with fewer interpretation al ambiguities.  
In addition, the use of the Atacama Large Millimeter Array (ALMA) at its lowest frequency
bandpass (90 GHz) will also stretch coverage of stellar flares, and probe the most energetic
flares, with $\approx$ 100-fold increase in sensitivity over currently existing mm arrays.  
The combination of these powerful new radio telescopes, in conjunction with 
optical and X-ray observations, should provide additional insight into the 
detailed heating processes operating in stellar flares.

\acknowledgments
This project presents the results of VLA project AO160.  RAO gratefully acknowledges
support from SAO grant GO1-2014C and STScI grant HST-GO-8880.  SLH and JCA acknowledge funding
from HST-GO-8613 from the Space Telescope Science Institute, operated by the Association of
Universities for Research in Astronomy, Inc. for NASA. RAO thanks Tim Bastian for fruitful
discussions and a careful reading of the paper.

\facility{HST(STIS)} 
\facility{Chandra(HETGS)}
\facility{VLA}
\facility{McDonald(2.1m)}
\facility{McDonald(2.7m)}

\clearpage



\begin{deluxetable}{lll}
\tablewidth{0pt}
\tablenum{1}
\tablecolumns{3}
\tablecaption{DECAY TIMESCALES FOR RADIO FLARE \label{tbl:raddecay}}
\tablehead{\colhead{Time Interval\tablenotemark{a}} & \colhead{X-band = 3.6 cm = 8.4 GHz} & \colhead{C-band = 6 cm = 4.9 GHz} \\
\colhead{(hr)} & \colhead{(s)} & \colhead{(s)} }
\startdata
8.51--8.54 & 860$\pm$70 & $\ldots$\tablenotemark{b} \\
8.54--8.90 & 1524$\pm$14 & 2728$\pm$78 \\
8.96--9.44 & 2250$\pm$45 & 3055$\pm$114 \\
9.51--9.98& 2546$\pm$115 & 3673$\pm$283 \\
10.04--10.52& 4604$\pm$472 & 5510$\pm$798 \\
10.59--11.06& 3250$\pm$229 & 10427$\pm$1880 \\
11.13--11.58& 4356$\pm$465 & 16040$\pm$4687 \\
\enddata
\tablenotetext{a}{UT hours on 20 September 2001}
\tablenotetext{b}{6 cm emission peaks $\approx$ 130 seconds following 3.6 cm peak}
\end{deluxetable}

\begin{deluxetable}{llllccll}
\tablewidth{0pt}
\tablenum{2}
\tablecolumns{8}
\tablecaption{ STRONG UV EMISSION LINES \label{tbl:uvtbl}}
\tablehead{\colhead{Ion} &\colhead{log$_{10}$ T$_{\rm form}$}& \colhead{$\lambda_{\rm lab}$} & 
\colhead{desig.\tablenotemark{a}} 
& \colhead{v} & \colhead{FWHM} & \colhead{$\zeta_{\rm turb}$} & \colhead{f $\times 10^{14}$} \\
\colhead{} &\colhead{(K)}& \colhead{(\AA)} & \colhead{} & \colhead{(km/s)}  &\colhead{(km/s)} & \colhead{(km/s)} & 
\colhead{(erg cm$^{-2}$ s$^{-1}$)}     }
\startdata
Si~III & 4.5 & 1206.510 & Qn&  -1$\pm$1 &      45$\pm$2 & 26.5 &  2.28$\pm$0.07 \\
 &  &  & F$_{1}$ & 14$\pm$6 &   34$\pm$15 &  19.6 & 1.82$\pm$0.67 \\
 	&    &	        & F$_{2}$ & -7$\pm$5&    66$\pm$12& 39.0 &  5.46$\pm$0.82 \\
N~V & 5.3 & 1238.821 & Qn &  -4.4$\pm$0.6 &   31$\pm$2 & $\ldots$ &  1.49$\pm$0.04 \\
    &     &          & Qb & -1$\pm$2 &       70$\pm$7 & 36.2 &  1.08$\pm$0.05 \\
   &  &  & F$_{1}$ &     5$\pm$4 &       34$\pm$9 & $\ldots$ &  1.66$\pm$0.37\\
    &     &          & F$_{2}$ &  -1$\pm$3 &    29$\pm$6 & $\ldots$ &  1.59$\pm$0.31 \\
Si~IV & 4.8 & 1393.755 & Qn &  -0.2$\pm$0.6 &     29$\pm$2 & 12.6 &  1.23$\pm$0.04 \\
      &     &          & Qb & 9$\pm$3 &    109$\pm$9& 64.2 &  1.08$\pm$0.08 \\
 &   &   & F$_{1}$&     2$\pm$1 &    25$\pm$3 & 8.9 &  4.92$\pm$0.45\\
      &     &         & F$_{2}$ &   -2$\pm$2 &     44$\pm$4 & 23.1 &  5.53$\pm$0.46 \\
Si~IV & 4.8 & 1402.773 & Qn &   -2.1$\pm$0.5 &  29$\pm$2 & 15.3 &  0.82$\pm$0.03 \\
      &     &          & Qb & 12$\pm$8 &    126$\pm$23 & 75 &   0.44$\pm$0.06 \\
 &  &  & F$_{1}$&   3$\pm$1  &     22$\pm$3 & 9.7 &  2.55$\pm$0.32 \\
      &     &         & F$_{2}$ &  -6$\pm$2 &   47$\pm$5 & 26.8 &  3.96$\pm$0.39 \\
C~IV & 5.0 & 1548.202 & Qn &   -0.4$\pm$0.4 &   34$\pm$1 & 11.7 &  0.54$\pm$0.10 \\
     &     &          & Qb & 1$\pm$2 &   135$\pm$10 & 79.5 &  2.72$\pm$0.20 \\
&  &  & F$_{1}$&    0$\pm$2 &      35$\pm$4  & 12.6 &  10.67$\pm$1.07 \\
    &      &         & F$_{2}$ &  -3$\pm$3 &     53$\pm$6 & 27.3 &  11.05$\pm$0.98 \\
C~IV & 5.0 & 1550.774 & Qn &  -3.4$\pm$0.7 &     31$\pm$2 & 8.0 &  2.64$\pm$0.09 \\
     &     &          & Qb & 10$\pm$6 &       91$\pm$16 & 51.9 & 1.18$\pm$0.15 \\
 &  &   & F$_{1}$ & -5$\pm$3 &   37$\pm$7 & 14.5 &  6.08$\pm$0.97 \\
     &     &          & F$_{2}$ & 0$\pm$4&   58$\pm$8 & 30.8 &  8.69$\pm$1.06 \\
\enddata
\tablenotetext{a}{Qn=narrow component of quiescence, Qb=broad component of quiescence \\
F$_{1}$=flare in 1st HST orbit; F$_{2}$=flare in 3rd HST orbit (see Figure~\ref{fig:timeline}). }
\end{deluxetable}

\begin{deluxetable}{llllllll}
\tablewidth{0pt}
\tablenum{3}
\tablecolumns{6}
\footnotesize
\tablecaption{ CHANDRA X-RAY FLARE PROPERTIES ON 20 SEPTEMBER 2001\label{tbl:chanflare}}
\tablehead{
\colhead{Flare \#\tablenotemark{a}} &\colhead{Time of peak} & \colhead{Peak luminosity\tablenotemark{b}} &  
 \colhead{$\Delta$t$_{\rm flare}$} & \colhead{E$_{\rm flare}$} & \colhead{$\tau_{d}$} \\
\colhead{} &\colhead{hh:mm (UT) } & \colhead{(10$^{28}$ erg s$^{-}$)}  & \colhead{(s)} &  \colhead{(10$^{31}$ erg)} &\colhead{(s)} }
\startdata
1 & 07:02& 7.7 &  5900 & 7.8 & 1730$\pm$170\\
2 &08:09& 28.7 &  900 & 0.21 & 112$\pm$26\\
3 &09:20 & 2.3 &  7796 & $\ldots$& 5180$\pm$3400\\
4 &10:19& 3.9 &  3167 & 5.6& 780$\pm$170 \\
5 & 12:27 & 3.4 &  4636 & $\ldots$ & 2850$\pm$600\\
6 &14:02 & 5.4 &  5473 & 3.4 & 950$\pm$260 \\
7 &16:43 & 16.6 &  1611 &$\ldots$ & 690$\pm$90\\
8 &17:17 & 12.7 &  8819  & 43. & 3800$\pm$780\\
9 &20:17 & 5.4 & 14130 & 20.& 4140$\pm$1210\\
\enddata
\tablenotetext{a}{Flare numbers are assigned in Figure~\ref{chanlc}.}
\tablenotetext{b}{Estimate of quiescent luminosity, 1.8$\times$10$^{28}$ erg s$^{-1}$, has been
subtracted.}
\end{deluxetable}

\begin{deluxetable}{lllllllll}
\tablewidth{0pt}
\tablenum{4}
\tablecolumns{9}
\footnotesize
\rotate
\tablecaption{ X-RAY FLARE SPECTRAL PROPERTIES \label{tbl:0thorder}}
\tablehead{ \colhead{Time\tablenotemark{a}} & \colhead{$\alpha$} & \colhead{kT$_{1}$} & \colhead{EM$_{1}$} & \colhead{kT$_{2}$}
& \colhead{EM$_{2}$} & \colhead{Z}  & \colhead{$\chi^{2}$ (dof)} \\
\colhead{}& \colhead{} & \colhead{ (MK)}  & \colhead{(10$^{51}$ cm$^{-3}$)} & \colhead{(MK)} & \colhead{(10$^{51}$
cm$^{-3}$)} & \colhead{} &\colhead{} }
\startdata
\footnotesize
Q & 0.233$_{-0.007}^{+0.019}$  & 4.3$_{-0.6}^{+0.1}$   & 5.4$_{-0.7}^{+1.6}$ & 9.2$_{-0.4}^{+0.4}$ & 5.6$_{-2.5}^{+0.6}$ & 0.24$_{-0.02}^{+0.01}$ & 124.1 (90)\\
Flare 1 & 0.53$_{-0.23}^{+0.26}$ & 3.9$_{-0.7}^{+1.5}$ & 2.9$_{-1.0}^{+1.2}$ & 
13.7$_{-1.4}^{+1.7}$  & 4.1$_{-1.7}^{+5.1}$ & 0.19$_{-0.07}^{0.3}$ & 26.7 (27)\\
Flare 4 & 0.35$_{-0.07}^{+0.29}$ & 4.3$_{-0.6}^{+1.0}$ & 4.9$_{-1.6}^{+1.6}$ & 
12.3$_{-1.2}^{+1.7}$ & 5.9$_{-3.6}^{+1.3}$ & 0.21 (0.12--0.32)& 28.9 (33)\\
Flare 5 & 1$_{-0.25}^{+0}$ & 3.6$_{-0.8}^{+0.8}$ & 2.6$_{-1.3}^{+1.9}$  & 
10.0$_{-1.8}^{+1.6}$  & 1.6$_{-0.9}^{+2.3}$  & 0.26$_{-0.15}^{+0.54}$ & 21.1 (15)\\
Flare 6 & 0.58$_{-0.27}^{+0.21}$  & 4.3$_{-0.6}^{+1.6}$  &4.7$_{-2.1}^{+3.4}$  & 
12.4$_{-1.7}^{+2.7}$  & 3.0$_{-1.7}^{+5.6}$ & 0.17$_{-0.1}^{+0.41}$  & 19.5 (23)\\
Flare 8 & 0.60$_{-0.09}^{+0.4}$ & 3.7$_{-0.5}^{+0.3}$  & 1.0$_{-0.4}^{+0.7}$  & 
30.0$_{-5.1}^{+3.4}$  & 2.6$_{-1.7}^{+1.2}$  & \tablenotemark{b} &47.5 (49)\\
Flare 9 & 0.45$_{_0.03}^{+0.13}$ & 4.5$_{-0.4}^{+0.1}$  & 4.2$_{-0.9}^{+1.2}$  & 
18.6$_{-1.6}^{+2.9}$  & 2.2$_{-0.9}^{+0.2}$ & \tablenotemark{c} & 87.1 (70)\\
\cline{1-8} \\
Flare1 Rise & 0.47$_{-0.35}^{+0.5}$ & 3.6$_{-0.6}^{+0.7}$ & 92.4$_{-92.4}^{+557.6}$ &
X\tablenotemark{e} & X\tablenotemark{e} & \tablenotemark{f} & 5.0 (4) \\
Flare1 Decay & 1$_{-0.35}^{+0}$  & 3.9\tablenotemark{d} & 2.9\tablenotemark{d} & 
16.9$_{-2.2}^{+3.7}$  & 67.8$_{-18.5}^{+15.4}$  & 0.16$_{-0.09}^{+0.25}$  & 18.7 (19) \\
Flare8 Rise & 0.51$_{-0.44}^{+0.49}$  &  35$_{-12}^{+109}$  & 10.2$_{-10.2}^{+6.7}$ &  
X\tablenotemark{e} & X\tablenotemark{e}& 1\tablenotemark{d} & 3.4 (3) \\
Flare8 Decay & 0.56$_{-0.02}^{+0.04}$ & 3.9$_{-0.7}^{+1.0}$ & 9.1$_{-3.8}^{+3.1}$  & 
13.8$_{-1.0}^{+1.5}$ & 11.3$_{-3.3}^{+1.7}$  & 0.17$_{-0.07}^{+0.09}$  & 70.2 (45) \\
Flare9 Rise & 0.26$_{-0.16}^{+0.45}$  & 6.8$_{-1.5}^{+1.1}$  & 15.1$_{-10.8}^{+203.6}$  
& X\tablenotemark{e} &
X\tablenotemark{e} &0.08$_{-0.06}^{+0.18}$  & 1.8 (5) \\
Flare9 Decay & 0.50$_{-0.04}^{+0.11}$ & 3.8$_{-0.3}^{+0.6}$  & 6.7$_{-1.8}^{+1.9}$  &  
11.4$_{-0.5}^{+0.8}$  & 6.5$_{-2.7}^{+0.8}$  &0.20$_{-0.05}^{+0.04}$ & 77.1 (64) \\
\enddata
\tablenotetext{a}{Time intervals shown in Figure~\ref{chanlc}.}
\tablenotetext{b}{Variable abundance APEC model fit w/following abundance constraints: 
O 0.82$_{-0.58}^{+0.68}$, Ne 4.63$_{-2.88}^{+2.29}$, Mg 4.92$_{-3.73}^{+6.48}$, 
Fe 0.94$_{-0.68}^{+0.97}$}
\tablenotetext{c}{Variable abundance APEC model fit w/following abundance constraints: 
O 0.37$_{-0.06}^{+0.17}$, Ne 1.06$_{-0.13}^{+0.32}$, Mg 0.77$_{-0.41}^{+0.36}$, 
Fe 0.28$_{-0.02}^{+0.10}$}
\tablenotetext{d}{Parameter fixed at value given}
\tablenotetext{e}{Parameter omitted from model fit}
\tablenotetext{f}{Variable abundance APEC model fit w/following abundance constraints:
Ne 0.49$_{-0.36}^{+1000}$, other abundances 0.08$_{-0.075}^{+9.22}$}
\end{deluxetable}

\begin{deluxetable}{lllll}
\tablewidth{0pt}
\tablenum{5}
\tablecolumns{5}
\tablecaption{STRONG EMISSION LINES IN X-RAY SPECTRA \label{tbl:xrayfluxes}}
\tablehead{
\colhead{Ion} & \colhead{Wavelength} & \colhead{Q} & \colhead{Large Flares\tablenotemark{a}} 
& \colhead{Small Flares\tablenotemark{a}} \\
\colhead{} & \colhead{(\AA)} & \multicolumn{3}{c}{--- Flux $\times$ 10$^{14}$ erg cm$^{-2}$ s$^{-1}$ ---} }
\startdata
Si~XIV & 6.18 & 2.16$\pm$0.64 &15.3$\pm$2.14 & 6.05$\pm$0.85 \\
Si~XIII & 6.65 & 4.25$\pm$0.74 & 12.2$\pm$1.79 & 9.84$\pm$0.92 \\
Mg~XII & 8.42 & 2.25$\pm$0.61 & 7.46$\pm$1.59 & 3.18$\pm$0.66 \\
Mg~XI & 9.17 & 2.98$\pm$0.54 & 4.62$\pm$1.28 & 3.62$\pm$0.58 \\
Ne~X & 12.13 &25.70$\pm$1.66 & 53.1$\pm$3.47 & 35.6$\pm$1.85 \\
Ne~IX & 13.45 & 22.8$\pm$1.77 & 23.9$\pm$3.10 & 23.9$\pm$1.87 \\
Fe~XVII & 15.01 &20.4$\pm$2.4 & 28.5$\pm$4.10 & 19.7$\pm$2.18 \\
O~VIII & 18.97 &64.2$\pm$4.59 & 73.7$\pm$7.82 & 65.7$\pm$4.29 \\
O~VII & 21.60 & 16.30$\pm$3.56 & 17.60$\pm$6.04 & 19.8$\pm$3.43 \\
\enddata
\tablenotetext{a}{``Large Flares'' = flares 1,2,7,8; ``Small Flares'' = flares 3--6,9
demarcated in Figure~\ref{chanlc}.}
\end{deluxetable}

\begin{deluxetable}{lccc}
\tablewidth{0pt}
\tablenum{6}
\tablecolumns{3}
\tablecaption{TEMPERATURE-SENSITIVE EMISSION LINE RATIOS IN X-RAY SPECTRA\label{tbl:tmpratios}}
\tablehead{
\colhead{Ratio\tablenotemark{a}} & \colhead{Q} & \colhead{Large Flares} & \colhead{Small Flares}  }
\startdata
Si~XIV 6.18/Si~XIII 6.45 & 0.51$\pm$0.18 &1.25$\pm$0.25 &0.62$\pm$0.10 \\
        & 10.3 MK &      14. MK & 11.1 MK \\
	&9.3$<$T$<$11.4 & 12.9$<$T$<$14.9 & 10.4$<$T$<$11.6 \\
Mg~XII 8.42/Mg~XI 9.17 & 0.76$\pm$0.25 & 1.61$\pm$0.56 & 0.88$\pm$0.23 \\
        & 7.4MK & 9.3MK & 7.7MK \\
	& 6.5$<$T$<$7.9 & 8.1$<$T$<$10.3 &7.0$<$T$<$8.2 \\
Ne~X 12.13/Ne~IX 13.45 & 1.13$\pm$0.11 & 2.22$\pm$0.32 & 1.49$\pm$0.14 \\
         & 4.7MK & 5.6MK & 5.2MK \\
	& 4.6$<$T$<$4.9 & 5.6$<$T$<$6.2 & 4.9$<$T$<$5.3 \\
O~VIII 18.97/O~VII 21.60 & 3.94 $\pm$0.90 & 4.19$\pm$1.50 & 3.32$\pm$0.61 \\
        & 3.7MK & 3.8MK & 3.5MK \\
	& 3.4$<$T$<$4.0 & 3.2$<$T$<$4.3 & 3.2$<$T$<$3.7 \\
\enddata
\tablenotetext{a}{Energy flux ratios}
\end{deluxetable}

\begin{deluxetable}{llllll}
\tablewidth{0pt}
\tablenum{7}
\tablecolumns{6}
\tablecaption{ABUNDANCE-SENSITIVE EMISSION LINE RATIOS IN X-RAY SPECTRA \label{tbl:abundratios}}
\tablehead{\colhead{Ratio\tablenotemark{a}} & \colhead{FIP type\tablenotemark{b}} & \colhead{APEC\tablenotemark{c}} & \colhead{Q} & \colhead{Large Flares} & \colhead{Small Flares} }
\startdata
Si~XIII 6.65/Mg~XII 8.42 & LL & 1.06 & 1.9$\pm$0.6 & 1.6$\pm$0.4 & 3.1$\pm$0.7 \\
Fe~XVII 15.01/Mg~XI 9.17 & LL & 10.5 & 6.84$\pm$1.47 & 6.17$\pm$1.92 & 5.44$\pm$1.06 \\
Mg~XI 9.17/Ne~X 12.13 & LH & 0.32 & 0.116$\pm$0.003 & 0.087$\pm$0.003 & 0.102$\pm$0.002 \\
Fe~XVII 15.01/Ne~X 12.13 & LH & 4.1 & 0.79$\pm$0.11 & 0.54$\pm$0.08 & 0.55$\pm$0.07 \\
Ne~IX 13.45/O~VIII 18.9 & HH & 0.19 & 0.36$\pm$0.04 & 0.32$\pm$0.05 & 0.36$\pm$0.04 \\
\enddata
\tablenotetext{a}{Energy flux ratios}
\tablenotetext{b}{L=low FIP ($<$10 eV), H=high FIP ($>$10 eV) }
\tablenotetext{c}{theoretical value using abundances of \citet{gs1998,gs1999}. }
\end{deluxetable}


\clearpage

\end{document}